\documentclass[aps,prd,twoside,twocolumn,nofootinbib,10pt,showpacs,floatfix]{revtex4-1}
\usepackage{amsmath,amssymb}
\usepackage{txfonts}
\usepackage{graphicx,bm}
\usepackage{slashed}
\usepackage{epstopdf}
\usepackage{ulem} 
\usepackage[usenames]{color}
\usepackage{float}
\usepackage{subfigure}
\usepackage{subfigure}
\usepackage{rotating}
\usepackage{color}
\usepackage{multirow}
\usepackage{dcolumn}
\usepackage{overpic}
\usepackage{booktabs}
\usepackage{makecell}
\usepackage{diagbox}

\renewcommand\sout{\bgroup \color{red} \ULdepth=-.5ex \ULset}

\newsavebox{\tablebox}
\usepackage[colorlinks,
            citecolor=blue,
            anchorcolor=red,
            menucolor=red,
            linkcolor=red,
            filecolor=red,
            runcolor=red,
            urlcolor=blue,
            frenchlinks=red]{hyperref}
\begin{document}

\title{Exploring the electromagnetic properties of the $\Xi_c^{(\prime,\,*)} \bar D_s^*$ and $\Omega_c^{(*)} \bar D_s^*$ molecular states}

\author{Fu-Lai Wang$^{1,2,3}$}
\email{wangfl2016@lzu.edu.cn}
\author{Si-Qiang Luo$^{1,2,3}$}
\email{luosq15@lzu.edu.cn}
\author{Hong-Yan Zhou$^{1,2}$}
\email{zhouhy20@lzu.edu.cn}
\author{Zhan-Wei Liu$^{1,2,3}$}
\email{liuzhanwei@lzu.edu.cn}
\author{Xiang Liu$^{1,2,3,4}$}
\email{xiangliu@lzu.edu.cn}
\affiliation{$^1$School of Physical Science and Technology, Lanzhou University, Lanzhou 730000, China\\
$^2$Research Center for Hadron and CSR Physics, Lanzhou University and Institute of Modern Physics of CAS, Lanzhou 730000, China\\
$^3$Lanzhou Center for Theoretical Physics, Key Laboratory of Theoretical Physics of Gansu Province, and Frontiers Science Center for Rare Isotopes, Lanzhou University, Lanzhou 730000, China\\
$^4$Key Laboratory of Quantum Theory and Applications of MoE, Lanzhou University, Lanzhou 730000, China}

\begin{abstract}
This paper presents a systematic investigation of the electromagnetic properties of the hidden-charm molecular pentaquarks within the constituent quark model. Specifically, it focuses on two types of pentaquarks: the $\Xi_c^{(\prime,*)} \bar{D}_s^{*}$-type pentaquarks with double strangeness and the $\Omega_{c}^{(*)}\bar D_s^{*}$-type pentaquarks with triple strangeness. The study explores various electromagnetic properties, including the magnetic moments, the transition magnetic moments, and the radiative decay behavior of these pentaquarks. To ensure realistic calculations, the $S$-$D$ wave mixing effect and the coupled channel effect are taken into account. By examining the electromagnetic properties of the hidden-charm molecular pentaquarks with double and triple strangeness, this research contributes to
the deeper understanding of their spectroscopic behavior. These findings form a valuable addition to the ongoing investigation into the broader spectrum of properties exhibited by the hidden-charm molecular pentaquarks.
\end{abstract}
\maketitle

\section{Introduction}\label{sec1}

Since the observation of the charmoniumlike state $X(3872)$ in 2003, numerous new hadronic states have been experimentally observed, leading to extensive discussions about their properties. These efforts have significantly enriched our understanding of the hadron spectroscopy \cite{Liu:2013waa,Hosaka:2016pey,Chen:2016qju,Richard:2016eis,Lebed:2016hpi,Olsen:2017bmm,Guo:2017jvc,Brambilla:2019esw,Liu:2019zoy,Chen:2022asf,Meng:2022ozq,Amsler:2004ps,Swanson:2006st,
Godfrey:2008nc,Yamaguchi:2019vea,Albuquerque:2018jkn,Yuan:2018inv,Ali:2017jda,Dong:2017gaw,Faccini:2012pj,Drenska:2010kg,Pakhlova:2010zza}. Moreover, these studies have been valuable in deepening our understanding of the nonperturbative behavior of the strong interactions.

Among the various assignments proposed for the observed new hadronic states, the molecular state explanation has gained popularity. Notably, in 2015, the LHCb Collaboration observed two $P_c$ states \cite{Aaij:2015tga} and subsequently reported two substructures, namely $P_{c}(4440)$ and $P_{c}(4457)$ \cite{Aaij:2019vzc}, corresponding to the previously observed $P_c(4450)$ \cite{Aaij:2015tga}. Furthermore, they discovered a new state, $P_c(4312)$, through a detailed analysis of the $\Lambda_b\to J/\psi p K$ process \cite{Aaij:2019vzc}. The LHCb experiment has provided strong experimental evidence for the existence of the hidden-charm molecular pentaquarks of the $\Sigma_c \bar D^{(*)}$ type \cite{Wu:2010jy,Wang:2011rga,Yang:2011wz,Wu:2012md,Li:2014gra,Karliner:2015ina,Chen:2015loa}. In subsequent years, LHCb reported the evidence for $P_{cs}(4459)$ \cite{LHCb:2020jpq} and observed $P_{\psi s}^{\Lambda}(4338)$ \cite{LHCb:2022jad}. These exciting experimental advancements have not only enriched the hidden-charm pentaquark family \cite{Hofmann:2005sw,Wu:2010vk,Anisovich:2015zqa,Wang:2015wsa,Feijoo:2015kts,Chen:2015sxa,Chen:2016ryt,Lu:2016roh,Xiao:2019gjd,Shen:2020gpw,Zhang:2020cdi,Wang:2019nvm,Weng:2019ynv,Chen:2020uif,Peng:2020hql,Chen:2020opr,Liu:2020hcv,Dong:2021juy,Chen:2022onm,Chen:2020kco,Chen:2021cfl,Chen:2021spf,Du:2021bgb,Hu:2021nvs,Xiao:2021rgp,Zhu:2021lhd,Wang:2022neq,Wang:2022mxy,Karliner:2022erb,Yan:2022wuz,Meng:2022wgl,Azizi:2021utt,Chen:2021tip,Clymton:2021thh,Zou:2021sha,Lu:2021irg,Ferretti:2021zis,Paryev:2022zdx,Nakamura:2022jpd,Giachino:2022pws,Marse-Valera:2022khy,Ortega:2022uyu,Chen:2022wkh,Zhu:2022wpi,Yang:2022ezl,Garcilazo:2022edi,Feijoo:2022rxf}, but have also inspired theorists to investigate the hidden-charm molecular pentaquarks of the $P_{css(s)}$ type \cite{Wang:2020bjt,Wang:2021hql,Azizi:2021pbh,Azizi:2022qll}. In recent years, the Lanzhou group has conducted extensive studies on the mass spectra of the hidden-charm molecular pentaquarks with double and triple strangeness. Specifically, their investigations have focused on the $\Xi_c^{(\prime,*)} \bar{D}_s^{(*)}$ \cite{Wang:2020bjt} and $\Omega_{c}^{(*)}\bar D_s^{(*)}$ \cite{Wang:2021hql} interactions. These studies have provided valuable insights into the properties and characteristics of these exotic hadronic states.

Currently, the investigation of the properties of the hidden-charm molecular pentaquarks remains a fascinating and significant research topic in hadron physics. It offers valuable insights for constructing a comprehensive family of the hidden-charm molecular pentaquarks. The study of the electromagnetic properties serves as an effective approach to unveil the inner structures of hadrons. A notable example is the successful application of the constituent quark model in describing the magnetic moments of the decuplet and octet baryons \cite{Schlumpf:1993rm,Kumar:2005ei,Ramalho:2009gk}, with corresponding experimental data available \cite{Workman:2022ynf}.

Given the importance of the electromagnetic properties, it is crucial to investigate the electromagnetic characteristics of the hidden-charm molecular pentaquarks. Some discussions on the electromagnetic properties of the $\Sigma_c^{(*)} \bar{D}^{(*)}$-type and $\Xi_c^{(\prime,*)} \bar{D}^{(*)}$-type hidden-charm molecular pentaquarks have been conducted within the constituent quark model \cite{Wang:2016dzu,Gao:2021hmv,Li:2021ryu,Wang:2022tib}. These studies shed light on the inner structures of the discussed hidden-charm molecular pentaquarks. However, it is important to note that the exploration of the electromagnetic properties for the hidden-charm molecular pentaquarks is still in its early stages. Thus, further efforts are required to obtain a comprehensive understanding of the electromagnetic properties of various types of hidden-charm molecular pentaquarks.

In this study, our focus is on investigating the electromagnetic properties of the hidden-charm molecular pentaquarks with double strangeness, specifically the $\Xi_c^{(\prime,*)} \bar{D}_s^{*}$-type pentaquarks, as well as those with triple strangeness, namely the $\Omega_{c}^{(*)}\bar D_s^{*}$-type pentaquarks. These particular pentaquark states were initially predicted in Refs. \cite{Wang:2020bjt,Wang:2021hql}. Within the framework of the constituent quark model, we examine their magnetic moments, transition magnetic moments, and radiative decay behavior. Our realistic calculations incorporate the effects of the $S$-$D$ wave mixing and coupled channels. By undertaking this investigation, we aim to enhance our understanding of the electromagnetic properties of the hidden-charm molecular pentaquarks with double and triple strangeness, thereby contributing to the comprehensive knowledge of these intriguing exotic hadrons \cite{Wang:2020bjt,Wang:2021hql}.

The structure of this paper is as follows. In Sec. \ref{sec2}, we provide a detailed explanation of the methodology employed for calculating the electromagnetic properties of the hadronic molecules. Additionally, we present the electromagnetic properties of the $\Xi_c^{(\prime,*)} \bar{D}_s^{*}$ molecular states. In Sec. \ref{sec3}, we shift our focus to the electromagnetic properties of the $\Omega_{c}^{(*)}\bar D_s^{*}$ molecular states. Finally, we offer a concise summary of our findings in Sec. \ref{sec4}.

\section{The electromagnetic properties of the  $\Xi_c^{(\prime,*)} \bar{D}_s^{*}$ molecules}\label{sec2}

In this section, we thoroughly investigate the electromagnetic properties of two molecular states: the $\Xi_{c}^{\prime}\bar D_s^*$ state with $I(J^P)=1/2({3}/{2}^{-})$ and the $\Xi_{c}^{*}\bar D_s^*$ state with $I(J^P)=1/2({5}/{2}^{-})$ \cite{Wang:2020bjt}. Specifically, we analyze their magnetic moments, transition magnetic moments, and radiative decay behavior. These investigations yield valuable insights into the inner
structures of these states, offering significant information in this regard.

\subsection{The magnetic moments and the transition magnetic moments of the  $\Xi_c^{(\prime,*)} \bar{D}_s^{*}$ molecules}

In the context of the constituent quark model, the hadronic magnetic moment encompasses two key components: the spin magnetic moment and the orbital magnetic moment. Specifically, when considering the $z$-component of the spin magnetic moment operator for a given hadron, denoted as $\hat{\mu}_z^{\rm spin}$, it can be mathematically represented as follows \cite{Liu:2003ab,Huang:2004tn,Zhu:2004xa,Haghpayma:2006hu,Wang:2016dzu,Deng:2021gnb,Gao:2021hmv,Li:2021ryu,Zhou:2022gra,Wang:2022tib,Li:2021ryu,Schlumpf:1992vq,Schlumpf:1993rm,Cheng:1997kr,Ha:1998gf,Ramalho:2009gk,Girdhar:2015gsa,Menapara:2022ksj,Mutuk:2021epz,Menapara:2021vug,Menapara:2021dzi,Gandhi:2018lez,Dahiya:2018ahb,Kaur:2016kan,Thakkar:2016sog,Shah:2016vmd,Dhir:2013nka,Sharma:2012jqz,Majethiya:2011ry,Sharma:2010vv,Dhir:2009ax,Simonis:2018rld,Ghalenovi:2014swa,Kumar:2005ei,Rahmani:2020pol,Hazra:2021lpa,Gandhi:2019bju,Majethiya:2009vx,Shah:2016nxi,Shah:2018bnr,Ghalenovi:2018fxh}:
\begin{eqnarray}
 \hat{\mu}_z^{\rm spin}&=&\sum_{j}\frac{e_j}{2M_j}\hat{\sigma}_{zj},
\end{eqnarray}
where $e_j$, $M_j$, and $\hat{\sigma}_{zj}$ denote the charge, the mass, and the $z$-component of the Pauli spin operator of the $j$th constituent of the hadron, respectively. When examining the hadronic molecule comprised of a baryon and a meson, the $z$-component of the orbital magnetic moment operator, denoted as $\hat{\mu}_z^{\rm orbital}$, can be expressed in the following manner \cite{Cheng:1997kr,Liu:2003ab,Huang:2004tn,Haghpayma:2006hu,Sharma:2010vv,Sharma:2012jqz,Girdhar:2015gsa,Wang:2016dzu,Dahiya:2018ahb,Gao:2021hmv,Li:2021ryu,Zhou:2022gra,Wang:2022tib}:
\begin{eqnarray}
 \hat{\mu}_z^{\rm orbital}&=&\mu_{bm}^L\hat{L}_z\nonumber\\
  &=&\left(\frac{M_{m}}{M_{b}+M_{m}}\frac{e_b}{2M_b}+\frac{M_{b}}{M_{b}+M_{m}}\frac{e_m}{2M_m}\right)\hat{L}_z,
\end{eqnarray}
where the subscript $b$ corresponds to the baryon, while the subscript $m$ pertains to the meson. Furthermore, $\hat{L}_z$ denotes the $z$-component of the orbital angular momentum operator linking the baryon and the meson. In this study, the masses of the $S$-wave charmed baryons and the $S$-wave charmed-strange meson are extracted from the Particle Data Group \cite{Workman:2022ynf} for reference.

As extensively discussed in various references such as \cite{Liu:2003ab,Huang:2004tn,Zhu:2004xa,Haghpayma:2006hu,Wang:2016dzu,Gao:2021hmv,Li:2021ryu,Deng:2021gnb,Zhou:2022gra,Wang:2022tib,Schlumpf:1992vq,Schlumpf:1993rm,Cheng:1997kr,Ha:1998gf,Ramalho:2009gk,Girdhar:2015gsa,Menapara:2022ksj,Mutuk:2021epz,Menapara:2021vug,Menapara:2021dzi,Gandhi:2018lez,Dahiya:2018ahb,Kaur:2016kan,Thakkar:2016sog,Shah:2016vmd,Dhir:2013nka,Sharma:2012jqz,Majethiya:2011ry,Sharma:2010vv,Dhir:2009ax,Simonis:2018rld,Ghalenovi:2014swa,Kumar:2005ei,Gandhi:2019bju,Rahmani:2020pol,Hazra:2021lpa,Majethiya:2009vx,Shah:2016nxi,Shah:2018bnr,Ghalenovi:2018fxh}, the magnetic moments of the hadrons ($\mu_{H}$) and the transition magnetic moments between the hadrons ($\mu_{H \to H^{\prime}}$) are frequently estimated by evaluating the expectation values of the $z$-component of the total magnetic moment operator ($\hat{\mu}_z$), which can be represented as
\begin{eqnarray}
\mu_{H}&=&\left\langle{J_{H},J_{H}|\hat\mu_{z}|J_{H},J_{H}}\right\rangle,\label{magneticmoment}\\
\mu_{H \to H^{\prime}}&=&\left\langle{J_{H^{\prime}},J_{z}|\hat\mu_{z}|J_{H},J_{z}}\right\rangle^{J_z={\rm Min}\{J_H,\,J_{H^{\prime}}\}}.\label{transitionmagneticmoment}
\end{eqnarray}
Here, $\hat{\mu}_z=\hat{\mu}_z^{\rm spin}+\hat{\mu}_z^{\rm orbital}$, and $H^{(\prime)}$ stands for either the fundamental hadron or the compound hadron. In the realistic calculations, the previous theoretical studies commonly employ the maximum value of the third component of the total angular momentum quantum number for the hadron to determine the hadronic magnetic moment. Similarly, they consider the maximum third component of the total angular momentum quantum number of the lowest state of the total angular momentum to discuss the transition magnetic moment between the hadrons \cite{Liu:2003ab,Huang:2004tn,Zhu:2004xa,Haghpayma:2006hu,Wang:2016dzu,Gao:2021hmv,Li:2021ryu,Deng:2021gnb,Zhou:2022gra,Wang:2022tib,Schlumpf:1992vq,Schlumpf:1993rm,Cheng:1997kr,Ha:1998gf,Ramalho:2009gk,Girdhar:2015gsa,Menapara:2022ksj,Mutuk:2021epz,Menapara:2021vug,Menapara:2021dzi,Gandhi:2018lez,Dahiya:2018ahb,Kaur:2016kan,Thakkar:2016sog,Shah:2016vmd,Dhir:2013nka,Sharma:2012jqz,Majethiya:2011ry,Sharma:2010vv,Dhir:2009ax,Simonis:2018rld,Ghalenovi:2014swa,Kumar:2005ei,Gandhi:2019bju,Rahmani:2020pol,Hazra:2021lpa,Majethiya:2009vx,Shah:2016nxi,Shah:2018bnr,Ghalenovi:2018fxh}. In our current study, we adopt the same model and convention as previous theoretical works for calculating the hadronic magnetic moments and the hadronic transition magnetic moments \cite{Liu:2003ab,Huang:2004tn,Zhu:2004xa,Haghpayma:2006hu,Wang:2016dzu,Gao:2021hmv,Li:2021ryu,Deng:2021gnb,Zhou:2022gra,Wang:2022tib,Schlumpf:1992vq,Schlumpf:1993rm,Cheng:1997kr,Ha:1998gf,Ramalho:2009gk,Girdhar:2015gsa,Menapara:2022ksj,Mutuk:2021epz,Menapara:2021vug,Menapara:2021dzi,Gandhi:2018lez,Dahiya:2018ahb,Kaur:2016kan,Thakkar:2016sog,Shah:2016vmd,Dhir:2013nka,Sharma:2012jqz,Majethiya:2011ry,Sharma:2010vv,Dhir:2009ax,Simonis:2018rld,Ghalenovi:2014swa,Kumar:2005ei,Gandhi:2019bju,Rahmani:2020pol,Hazra:2021lpa,Majethiya:2009vx,Shah:2016nxi,Shah:2018bnr,Ghalenovi:2018fxh}. In order to provide a comprehensive analysis, it is necessary to discuss the wave functions of the hadronic states under consideration. These wave functions encompass various aspects, including the color part, the flavor part, the spin part, and the spatial part.
Regarding the color wave function, it is straightforwardly represented by the constant value 1, as the color aspect is typically treated uniformly in our context. On the other hand, the flavor-spin wave function can be constructed by taking into account the symmetry constraints imposed by the system. Finally, the spatial wave function can be derived by quantitatively studying the mass spectrum of the corresponding hadron \cite{Wang:2022tib}.

In the subsequent analysis, we delve into the magnetic moments and the transition magnetic moments of two specific molecular states: the $\Xi_{c}^{\prime}\bar D_s^*$ molecule with $I(J^P)=1/2({3}/{2}^{-})$ and the $\Xi_{c}^{*}\bar D_s^*$ molecular state with $I(J^P)=1/2({5}/{2}^{-})$. To accomplish this, we employ three distinct scenarios: the single-channel analysis, the $S$-$D$ wave mixing analysis, and the coupled channel analysis. These scenarios allow us to explore the influence of the $S$-$D$ wave mixing effect and the coupled channel effect on the magnetic moments and the transition magnetic moments of the $\Xi_c^{(\prime,*)} \bar{D}_s^{*}$ molecular states. By employing the aforementioned procedures, we can elucidate the respective contributions of these effects to the magnetic moments and the transition magnetic moments of the molecular states under investigation.

\subsubsection{The single channel analysis}

Firstly, we investigate the magnetic moments and the transition magnetic moments of the $\Xi_{c}^{\prime}\bar D_s^*$ molecular state with $I(J^P)={1}/{2}({3}/{2}^{-})$ and the $\Xi_{c}^{*}\bar D_s^*$ molecule with  $I(J^P)={1}/{2}({5}/{2}^{-})$, while considering only the $S$-wave component. The flavor wave functions of these states, denoted as $|I,I_3\rangle$, can be expressed as  \cite{Wang:2020bjt}
\begin{eqnarray*}
\left|\frac{1}{2},\frac{1}{2}\right\rangle=|\Xi_c^{(\prime,*)+}{D}_s^{*-}\rangle,~~~~~~~~
\left|\frac{1}{2},-\frac{1}{2}\right\rangle=|\Xi_c^{(\prime,*)0}{D}_s^{*-}\rangle,
\end{eqnarray*}
where $I$ and $I_3$ represent the isospins and the isospin third components of the $\Xi_c^{(\prime,*)}\bar{D}_s^{*}$ systems, respectively. Furthermore, the spin wave functions $|S,S_3\rangle$ for these states can be constructed using the following coupling scheme \cite{Wang:2020bjt}
\begin{eqnarray*}
\Xi_c^{\prime}\bar{D}_s^{*}:\,|S,S_3\rangle&=&\sum_{S_{\Xi_c^{\prime}},S_{\bar{D}_s^{*}}}C^{SS_3}_{\frac{1}{2}S_{\Xi_c^{\prime}},1S_{\bar{D}_s^{*}}}\left|\frac{1}{2},S_{\Xi_c^{\prime}}\right\rangle\left|1,S_{\bar{D}_s^{*}}\right\rangle,\\
\Xi_c^{*}\bar{D}_s^{*}:\,|S,S_3\rangle&=&\sum_{S_{\Xi_c^{*}},S_{\bar{D}_s^{*}}}C^{S S_3}_{\frac{3}{2}S_{\Xi_c^{*}},1S_{\bar{D}_s^{*}}}\left|\frac{3}{2},S_{\Xi_c^{*}}\right\rangle\left|1,S_{\bar{D}_s^{*}}\right\rangle.
\end{eqnarray*}
Here, $S$ and $S_3$ represent the total spins and the total spin third components for the $\Xi_c^{(\prime,*)}\bar{D}_s^{*}$ systems, respectively. The Clebsch-Gordan coefficient $C^{ef}_{ab,cd}$ is utilized in the coupling scheme. Additionally, $S_{\Xi_c^{\prime}}$, $S_{\Xi_c^{*}}$, and $S_{\bar{D}_s^{*}}$ correspond to the spin third components of $\Xi_c^{\prime}$, $\Xi_c^{*}$, and $\bar{D}_s^{*}$, respectively.

With the aforementioned setup, we can now proceed to calculate the magnetic moments of the $\Xi_{c}^{\prime}\bar D_s^*$ molecule with $I(J^P)={1}/{2}({3}/{2}^{-})$ and the $\Xi_{c}^{*}\bar D_s^*$ molecular state with $I(J^P)={1}/{2}({5}/{2}^{-})$, such as
\begin{eqnarray}
  \mu_{\Xi_c^{\prime}\bar D_s^*|{3}/{2}^-\rangle}^{I_3=1/2}&=&\left\langle \chi_{\Xi_c^{\prime+}{{D}_s^{*-}}}^{\left|\frac{1}{2},\frac{1}{2}\right\rangle|1,1\rangle} \right|\hat{\mu}_z\left| \chi_{\Xi_c^{\prime+}{{D}_s^{*-}}}^{\left|\frac{1}{2},\frac{1}{2}\right\rangle|1,1\rangle}  \right\rangle\nonumber\\
  &=&\mu_{\Xi_c^{\prime+}}+\mu_{D_s^{*-}}.
\end{eqnarray}
Here, $\chi_f^s$ represents the spin and flavor wave functions of the hadron, while the superscript $s$ indicates the spin wave function and the subscript $f$ denotes the flavor wave function. Furthermore, in the context of the single channel analysis of the hadronic magnetic moment, the overlap of the relevant spatial wave function is 1. For brevity, this factor is omitted in the above expression.

To determine the magnetic moments of the $\Xi_c^{\prime(*)}$ baryons and the $\bar D_s^*$ meson, we employ the constituent quark model. Initially, let us define the flavor and spin wave functions of these particles. The flavor wave functions can be expressed as follows:
\begin{eqnarray*}
\Xi_c^{\prime(*)+}:\frac{1}{\sqrt{2}}\left(usc+suc\right),~~
\Xi_c^{\prime(*)0}:\frac{1}{\sqrt{2}}\left(dsc+sdc\right),~~
D_s^{*-}:\bar c s,
\end{eqnarray*}
while their corresponding spin wave functions $|S,S_3\rangle$ can be expressed as
\begin{eqnarray*}
\Xi_c^{\prime}&:&\left\{
  \begin{array}{l}
    \left|\dfrac{1}{2},\dfrac{1}{2}\right\rangle=\dfrac{1}{\sqrt{6}}\left(2\uparrow\uparrow\downarrow-\downarrow\uparrow\uparrow-\uparrow\downarrow\uparrow\right)\\
    \left|\dfrac{1}{2},-\dfrac{1}{2}\right\rangle=\dfrac{1}{\sqrt{6}}\left(\downarrow\uparrow\downarrow+\uparrow\downarrow\downarrow-2\downarrow\downarrow\uparrow\right)
  \end{array}
\right.,\\
\Xi_c^{*}&:&\left\{
  \begin{array}{l}
    \left|\dfrac{3}{2},\dfrac{3}{2}\right\rangle=\uparrow\uparrow\uparrow\\
    \left|\dfrac{3}{2},\dfrac{1}{2}\right\rangle=\dfrac{1}{\sqrt{3}}\left(\downarrow\uparrow\uparrow+\uparrow\downarrow\uparrow+\uparrow\uparrow\downarrow\right)\\
     \left|\dfrac{3}{2},-\dfrac{1}{2}\right\rangle=\dfrac{1}{\sqrt{3}}\left(\downarrow\downarrow\uparrow+\uparrow\downarrow\downarrow+\downarrow\uparrow\downarrow\right)\\
    \left|\dfrac{3}{2},-\dfrac{3}{2}\right\rangle=\downarrow\downarrow\downarrow
  \end{array}
\right.,\\
\bar D^{*}_s&:&\left\{
  \begin{array}{l}
    \left|1,1\right\rangle=\uparrow\uparrow\\
    \left|1,0\right\rangle=\dfrac{1}{\sqrt{2}}\left(\uparrow\downarrow+\downarrow\uparrow\right)\\
    \left|1,-1\right\rangle=\downarrow\downarrow
  \end{array}
\right..
\end{eqnarray*}
Here, the notations $\uparrow$ and $\downarrow$ denote the third components of the quark spins, with values of $+{1}/{2}$ and $-{1}/{2}$, respectively.

Based on the flavor and spin wave functions of the $\Xi_c^{\prime(*)}$ baryons and the $\bar D_s^*$ meson, we can proceed to calculate their magnetic moments. As an example, let us deduce the magnetic moment of the $\Xi_c^{\prime +}$ baryon as follows:
\begin{eqnarray}
\mu_{\Xi^{\prime +}_c}&=&\left\langle \chi_{\frac{1}{\sqrt{2}}\left(usc+suc\right)}^{\frac{1}{\sqrt{6}}\left(2\uparrow\uparrow\downarrow-\downarrow\uparrow\uparrow-\uparrow\downarrow\uparrow\right)} \right|\hat{\mu}_z\left| \chi_{\frac{1}{\sqrt{2}}\left(usc+suc\right)}^{\frac{1}{\sqrt{6}}\left(2\uparrow\uparrow\downarrow-\downarrow\uparrow\uparrow-\uparrow\downarrow\uparrow\right)} \right\rangle\nonumber\\
&=&\frac{2}{3}\mu_u+\frac{2}{3}\mu_s-\frac{1}{3}\mu_c.
\end{eqnarray}
In this study, we adopt the following definition for the magnetic magneton of the quark: $\mu_{q}=-\mu_{\bar q}={e_q}/{2M_q}$, where $e_q$ represents the charge of the quark and $M_q$ denotes the constituent mass of the quark. Utilizing this definition, we can derive the expressions for the magnetic moments of the $\Xi_c^{\prime(*)}$ baryons and the $\bar D_s^*$ meson. For the numerical analysis, we utilize the constituent quark masses $M_{u}=0.336\,\mathrm{GeV}$, $M_{d}=0.336\,\mathrm{GeV}$, $M_{s}=0.450\,\mathrm{GeV}$, and $M_{c}=1.680\,\mathrm{GeV}$ to quantitatively investigate the electromagnetic properties of these discussed hadrons. These constituent quark masses are sourced from Ref. \cite{Kumar:2005ei} and are widely employed in studies related to the magnetic moments of the hadronic molecular states \cite{Li:2021ryu,Zhou:2022gra,Wang:2022tib}.

In Table~\ref{MT1}, we present the expressions and numerical results for the magnetic moments of the $\Xi_c^{\prime(*)}$ baryons and the $\bar D_s^*$ meson. Our obtained results align with those reported in previous works \cite{Kumar:2005ei,Sharma:2010vv,Glozman:1995xy,Patel:2007gx,Simonis:2018rld,Ghalenovi:2014swa,Zhang:2021yul,Aliev:2015axa}. In this study, the magnetic moments  and the transition magnetic moments of hadrons are expressed in units of the nuclear magneton $\mu_N = {e}/{2M_P}$ with $M_P = 0.938\,\mathrm{GeV}$ \cite{Workman:2022ynf}. As shown in Table~\ref{MT1}, the $\Xi_c^{\prime +}$ and $\Xi_c^{\prime 0}$ baryons exhibit distinct magnetic moments, while the magnetic moment of the $\Xi_c^{* +}$ differs from that of the $\Xi_c^{* 0}$. This discrepancy arises from the notable difference in the magnetic magnetons between the up quark and the down quark, namely, $\mu_u = 1.862\,\mu_N$ and $\mu_d = -0.931\,\mu_N$. Moreover, the $\Xi_c^{\prime 0}$ and $\Xi_c^{* 0}$ exhibit approximately equal magnetic moments.
\renewcommand\tabcolsep{0.09cm}
\renewcommand{\arraystretch}{1.80}
\begin{table}[!htbp]
  \caption{The magnetic moments and the transition magnetic moments of the $\Xi_c^{\prime(*)}$ baryons and the $\bar D^{*}_s$ meson. The magnetic moment and the transition magnetic moment are given in units of $\mu_N$, where $\mu_N$ denotes the nuclear magneton. The expressions for the magnetic moments and the transition magnetic moments are enclosed in the square brackets in the second column.}
  \label{MT1}
\begin{tabular}{c|l|l}
\toprule[1.0pt]
\toprule[1.0pt]
Quantities &  \multicolumn{1}{c|}{Our work}  &  \multicolumn{1}{c}{Other works} \\\hline
$\mu_{\Xi^{\prime +}_c}$                   & $0.654\,\left[\frac{2}{3}\mu_u+\frac{2}{3}\mu_s-\frac{1}{3}\mu_c\right]$   &  0.65 \cite{Glozman:1995xy},\,0.67 \cite{Zhang:2021yul} \\
$\mu_{\Xi^{\prime 0}_c}$                   & $-1.208\,\left[\frac{2}{3}\mu_d+\frac{2}{3}\mu_s-\frac{1}{3}\mu_c\right]$  & $-1.20$ \cite{Aliev:2015axa},\,$-1.20$ \cite{Zhang:2021yul}\\
$\mu_{\Xi^{*+}_c}$                         & $1.539\,\left[\mu_u+\mu_s+\mu_c\right]$                                    &1.51 \cite{Patel:2007gx},\,1.59 \cite{Sharma:2010vv}\\
$\mu_{\Xi^{*0}_c}$                         & $-1.254\,\left[\mu_d+\mu_s+\mu_c\right]$                                   &$-1.20$ \cite{Simonis:2018rld},\,$-1.18$ \cite{Ghalenovi:2014swa}\\
$\mu_{D^{*-}_s}$                           & $-1.067\,\left[\mu_{\overline{c}}+\mu_s\right]$                            &$-1.00$ \cite{Simonis:2018rld},\,$-1.08$ \cite{Zhang:2021yul}\\
$\mu_{\Xi^{*+}_c \to \Xi^{\prime +}_c}$    & $0.199\,\left[\frac{\sqrt{2}}{3}(\mu_u+\mu_s-2\mu_c)\right]$               &0.17 \cite{Kumar:2005ei},\,0.16 \cite{Aliev:2009jt}\\
$\mu_{\Xi^{*0}_c \to \Xi^{\prime 0}_c}$    & $-1.117\,\left[\frac{\sqrt{2}}{3}(\mu_d+\mu_s-2\mu_c)\right]$              &$-1.07$ \cite{Simonis:2018rld},\,$-1.03$ \cite{Simonis:2018rld}\\
\bottomrule[1.0pt]
\bottomrule[1.0pt]
\end{tabular}
\end{table}

Based on our obtained magnetic moments of the $\Xi_c^{\prime(*)}$ baryons and the $\bar D^{*}_s$ meson, we can get the numerical results of the magnetic moments of the $\Xi_{c}^{\prime}\bar D_s^*$ molecule with $I(J^P)=1/2({3}/{2}^{-})$ and the $\Xi_{c}^{*}\bar D_s^*$ molecular state with $I(J^P)=1/2({5}/{2}^{-})$. In Table~\ref{ME1}, we present the expressions and numerical results of the magnetic moments of the $\Xi_{c}^{\prime}\bar D_s^*$ molecular state with $I(J^P)=1/2({3}/{2}^{-})$ and the $\Xi_{c}^{*}\bar D_s^*$ molecular state with $I(J^P)=1/2({5}/{2}^{-})$ when performing the single channel analysis.
\renewcommand\tabcolsep{0.55cm}
\renewcommand{\arraystretch}{1.50}
\begin{table}[!htbp]
  \caption{The expressions and numerical results of the magnetic moments of the $\Xi_{c}^{\prime}\bar D_s^*$ molecular state with $I(J^P)=1/2({3}/{2}^{-})$ and the $\Xi_{c}^{*}\bar D_s^*$ molecular state with $I(J^P)=1/2({5}/{2}^{-})$ when only the $S$-wave component is considered.}\label{ME1}
\begin{tabular}{c|c|c}
\toprule[1.0pt]
\toprule[1.0pt]
Physical quantities &  Expressions & Values \\\hline
$\mu_{\Xi_{c}^{\prime}\bar D_s^*|{3}/{2}^-\rangle}^{I_3=1/2}$  & $\mu_{\Xi_{c}^{\prime+}}+\mu_{D_{s}^{*-}}$& $-0.414~\mu_N$ \\
$\mu_{\Xi_{c}^{\prime}\bar D_s^*|{3}/{2}^-\rangle}^{I_3=-1/2}$  & $\mu_{\Xi_{c}^{\prime0}}+\mu_{D_{s}^{*-}}$& $-2.275~\mu_N$\\
$\mu_{\Xi_{c}^{*}\bar D_s^*|{5}/{2}^-\rangle}^{I_3=1/2}$  & $\mu_{\Xi_{c}^{*+}}+\mu_{D_{s}^{*-}}$& $0.472~\mu_N$ \\
$\mu_{\Xi_{c}^{*}\bar D_s^*|{5}/{2}^-\rangle}^{I_3=-1/2}$  & $\mu_{\Xi_{c}^{*0}}+\mu_{D_{s}^{*-}}$& $-2.321~\mu_N$\\
\bottomrule[1.0pt]
\bottomrule[1.0pt]
\end{tabular}
\end{table}

As presented in Table~\ref{ME1}, the magnetic moments of the $\Xi_c^{\prime} \bar D_s^*|{3}/{2}^-\rangle$ molecule with $I_3={1}/{2}$, the $\Xi_c^{\prime} \bar D_s^*|{3}/{2}^-\rangle$ molecule with $I_3=-{1}/{2}$, the $\Xi_c^{*} \bar D_s^*|{5}/{2}^-\rangle$ molecule with $I_3={1}/{2}$, and the $\Xi_c^{*} \bar D_s^*|{5}/{2}^-\rangle$ molecule with $I_3=-{1}/{2}$ are $-0.414~\mu_N$, $-2.275~\mu_N$, $0.472~\mu_N$, and $-2.321~\mu_N$, respectively. Notably, the magnetic moment of the $\Xi_c^{\prime} \bar D_s^*|{3}/{2}^-\rangle$ molecular state can be obtained as the sum of the magnetic moments of the $\Xi_c^{\prime}$ baryon and the $\bar D_s^*$ meson. Furthermore, the magnetic moment of the $\Xi^{\prime +}_c$ significantly differs from that of the $\Xi^{\prime 0}_c$, resulting in a distinct magnetic moment for the $\Xi_c^{\prime} \bar D_s^*|{3}/{2}^-\rangle$ molecule with $I_3={1}/{2}$ compared to that with $I_3=-{1}/{2}$. Similarly, the $\Xi_c^{*} \bar D_s^*|{5}/{2}^-\rangle$ molecular state exhibits different magnetic moments for various $I_3$ quantum numbers. In addition, the magnetic moments of the $\Xi_c^{\prime} \bar D_s^*|{3}/{2}^-\rangle$ molecule with $I_3=-{1}/{2}$ and the $\Xi_c^{*} \bar D_s^*|{5}/{2}^-\rangle$ molecule with $I_3=-{1}/{2}$ are nearly the same, owing to the close magnetic moments of the $\Xi_c^{\prime 0}$ and $\Xi_c^{* 0}$.

In addition to investigating the magnetic moments, we also examine the transition magnetic moments between the $\Xi_{c}^{\prime}\bar D_s^*$ molecular state with $I(J^P)=1/2({3}/{2}^{-})$ and the $\Xi_{c}^{*}\bar D_s^*$ molecular state with $I(J^P)=1/2({5}/{2}^{-})$. The transition magnetic moment between these two states can be determined using the following expression:
\begin{eqnarray}
&&\mu_{\Xi_c^{*}\bar D_s^*|{5}/{2}^-\rangle \to \Xi_c^{\prime}\bar D_s^*|{3}/{2}^-\rangle}^{I_3=1/2}\nonumber\\
&&=\left\langle\chi_{\Xi_c^{\prime+}{D}_s^{*-}}^{\left|\frac{1}{2},\frac{1}{2}\right\rangle|1,1\rangle} \right|\hat{\mu}_z\left| \chi_{\Xi_c^{*+}{D}_s^{*-}}^{\sqrt{\frac{2}{5}}\left|\frac{3}{2},\frac{3}{2}\right\rangle\left|1,0\right\rangle+\sqrt{\frac{3}{5}}\left|\frac{3}{2},\frac{1}{2}\right\rangle\left|1,1\right\rangle}\right\rangle\nonumber\\
&&=\sqrt{\frac{3}{5}}\mu_{\Xi^{*+}_c \to \Xi^{\prime +}_c}.
\end{eqnarray}
Hence, the transition magnetic moment for the $\Xi_c^{*}\bar D_s^*|{5}/{2}^-\rangle \to \Xi_c^{\prime}\bar D_s^*|{3}/{2}^-\rangle \gamma$ process can be connected to that of the $\Xi^{*}_c \to \Xi^{\prime }_c \gamma$ process. It should be noted that the spatial wave functions of the initial and final states may influence the transition magnetic moment, and this aspect will be addressed in the subsequent subsection. Next, we proceed to estimate the transition magnetic moment for the $\Xi^{*+}_c \to \Xi^{\prime +}_c \gamma$ process, which can be obtained from the expression
\begin{eqnarray}
\mu_{\Xi^{*+}_c \to \Xi^{\prime +}_c}&=&\left\langle \chi_{\frac{1}{\sqrt{2}}\left(usc+suc\right)}^{\frac{1}{\sqrt{3}}\left(\downarrow\downarrow\uparrow+\uparrow\downarrow\downarrow+\downarrow\uparrow\downarrow\right)} \right|\hat{\mu}_z\left| \chi_{\frac{1}{\sqrt{2}}\left(usc+suc\right)}^{\frac{1}{\sqrt{6}}\left(2\uparrow\uparrow\downarrow-\downarrow\uparrow\uparrow-\uparrow\downarrow\uparrow\right)}\right\rangle\nonumber\\
&=&\frac{\sqrt{2}}{3}\left(\mu_u+\mu_s-2\mu_c\right).
\end{eqnarray}
Table~\ref{MT1} presents the expressions and numerical values of the transition magnetic moments for the $\Xi^{*+}_c \to \Xi^{\prime +}_c \gamma$ and $\Xi^{*0}_c \to \Xi^{\prime 0}_c\gamma$ processes. The obtained results from our analysis are in good agreement with the theoretical predictions reported in Refs. \cite{Kumar:2005ei,Aliev:2009jt,Simonis:2018rld}.

Based on the calculated transition magnetic moments for the $\Xi^{*+}_c \to \Xi^{\prime +}_c \gamma$ and $\Xi^{*0}_c \to \Xi^{\prime 0}_c\gamma$ processes, we can determine the values of the transition magnetic moments between the $\Xi_{c}^{\prime}\bar D_s^*$ molecule with $I(J^P)=1/2({3}/{2}^{-})$ and the $\Xi_{c}^{*}\bar D_s^*$ molecular state with $I(J^P)=1/2({5}/{2}^{-})$. Specifically, we find that
\begin{eqnarray*}
\mu_{\Xi_c^{*}\bar D_s^*|{5}/{2}^-\rangle \to \Xi_c^{\prime}\bar D_s^*|{3}/{2}^-\rangle}^{I_3=1/2}&=&0.154~\mu_N,\\
\mu_{\Xi_c^{*}\bar D_s^*|{5}/{2}^-\rangle \to \Xi_c^{\prime}\bar D_s^*|{3}/{2}^-\rangle}^{I_3=-1/2}&=&-0.866~\mu_N.
\end{eqnarray*}
It should be noted that the magnitude of the transition magnetic moment for the $\Xi^{*0}_c \to \Xi^{\prime 0}_c$ process is significantly larger than that for the $\Xi^{*+}_c \to \Xi^{\prime +}_c$ process \cite{Simonis:2018rld,Aliev:2009jt}. Consequently, the absolute value of $\mu_{\Xi_c^{*}\bar D_s^*|{5}/{2}^-\rangle \to \Xi_c^{\prime}\bar D_s^*|{3}/{2}^-\rangle}$ with $I_3=-1/2$ is considerably greater than that with $I_3=1/2$.

\subsubsection{The $S$-$D$ wave mixing analysis}

And then, we conduct further investigations on the magnetic moments and the transition magnetic moments of the $\Xi_{c}^{\prime}\bar D_s^*$ molecular state with $I(J^P)=1/2({3}/{2}^{-})$ and the $\Xi_{c}^{*}\bar D_s^*$ molecule with $I(J^P)=1/2({5}/{2}^{-})$ by considering the additional contribution from the $D$-wave channels. Our calculations encompass the following $S$-wave and $D$-wave channels for the $\Xi_{c}^{\prime}\bar D_s^*$ molecular state with $I(J^P)=1/2({3}/{2}^{-})$ and the $\Xi_{c}^{*}\bar D_s^*$ molecular state with $I(J^P)=1/2({5}/{2}^{-})$ \cite{Wang:2020bjt}
\begin{eqnarray*}
&&\Xi_c^{\prime}\bar D_s^{*}|{3}/{2}^-\rangle:~~|^{4} S_{3/2}\rangle,\,|^{2} D_{3/2}\rangle,\,|^{4} D_{3/2}\rangle,\nonumber\\
&&\Xi_c^{*}\bar D_s^{*}|{5}/{2}^-\rangle:~~|^{6} S_{5/2}\rangle,\,|^{2} D_{5/2}\rangle,\,|^{4} D_{5/2}\rangle,\,|^{6} D_{5/2}\rangle.
\end{eqnarray*}
Here, we adopt the notation $|^{2S+1} L_{J}\rangle$ to denote the spin $S$, orbital angular momentum $L$, and total angular momentum $J$ of the molecular state under consideration.

By considering the influence of the $S$-$D$ wave mixing effect, we can derive the magnetic moment and the transition magnetic moment of the molecular states through the following deductions
\begin{eqnarray}
&&\sum_{i,\,j} \mu_{\mathcal{A}_i \to \mathcal{A}_j}\langle \phi_{\mathcal{A}_j}|\phi_{\mathcal{A}_i}\rangle,\label{MSD1}\\
&&\sum_{i,\,j} \mu_{\mathcal{B}_i \to \mathcal{A}_j}\langle \phi_{\mathcal{A}_j}|\phi_{\mathcal{B}_i}\rangle,\label{MSD2}
\end{eqnarray}
respectively. In this context, $\mathcal{A}$ and $\mathcal{B}$ denote the two molecular states under discussion, while $\phi_{i}$ represents the spatial wave function of the respective $i$th channel.

When incorporating the contribution of the $D$-wave channels to analyze the electromagnetic properties of the molecular states, it is necessary to outline the procedure for determining the magnetic moments and the transition magnetic moments of the $D$-wave channels. In the case of these specific $D$-wave channels, their spin-orbital wave functions $|{ }^{2 S+1} L_{J}\rangle$ can be constructed by coupling the spin wave function $\left|S, m_{S}\right\rangle$ with the orbital wave function $Y_{L, m_{L}}$. Explicitly, they can be expressed as
\begin{eqnarray}
\left|{ }^{2 S+1} L_{J}\right\rangle=\sum_{m_{S}, m_{L}} C_{S m_{S}, L m_{L}}^{J M} \left|S, m_{S}\right\rangle Y_{L, m_{L}}.
\end{eqnarray}
Thus, the magnetic moment of the $\Xi_c^{\prime}\bar D^{*}|^{2} D_{3/2}\rangle$ channel with $I_3=1/2$ is given by
\begin{eqnarray}
\mu_{\Xi_c^{\prime}\bar D^{*}|^{2} D_{3/2}\rangle}^{I_3=1/2}&=&\frac{1}{5}\left(-\frac{1}{3} \mu_{\Xi_c^{\prime+}}+\frac{2}{3}\mu_{D_s^{*-}}+\mu_{\Xi_c^{\prime+}D_s^{*-}}^L\right)\nonumber\\
&&+\frac{4}{5}\left(\frac{1}{3} \mu_{\Xi_c^{\prime+}}-\frac{2}{3}\mu_{D_s^{*-}}+2\mu_{\Xi_c^{\prime+} D_s^{*-}}^L\right)\nonumber\\
&=&\frac{1}{5} \mu_{\Xi_c^{\prime+}}-\frac{2}{5}\mu_{D_s^{*-}}+\frac{9}{5}\mu_{\Xi_c^{\prime+} D_s^{*-}}^L,
\end{eqnarray}
and the transition magnetic moment of the $\Xi_c^{\prime}\bar D^{*}|^{4} D_{3/2}\rangle \to \Xi_c^{\prime}\bar D^{*}|^{2} D_{3/2}\rangle$ process with $I_3=1/2$ can be determined as follows
\begin{eqnarray}
\mu_{\Xi_c^{\prime}\bar D^{*}|^{4} D_{3/2}\rangle \to \Xi_c^{\prime}\bar D^{*}|^{2} D_{3/2}\rangle}^{I_3=1/2}
&=&-\frac{\sqrt{2}}{5}\left(\frac{2\sqrt{2}}{3} \mu_{\Xi_c^{\prime+}}-\frac{\sqrt{2}}{3}\mu_{D_s^{*-}}\right)\nonumber\\
&&-\frac{2\sqrt{2}}{5}\left(\frac{2\sqrt{2}}{3} \mu_{\Xi_c^{\prime+}}-\frac{\sqrt{2}}{3}\mu_{D_s^{*-}}\right)\nonumber\\
&=&-\frac{4}{5} \mu_{\Xi_c^{\prime+}}+\frac{2}{5}\mu_{D_s^{*-}}.
\end{eqnarray}
By employing the aforementioned procedure, we can derive the magnetic moments and the transition magnetic moments of the $D$-wave channels included in our calculation.

By referring to Eqs. (\ref{MSD1})-(\ref{MSD2}), the magnetic moments and the transition magnetic moments of the hadronic molecules rely on the relevant mixing channel components $\langle \phi_{\mathcal{A}_j}|\phi_{\mathcal{A}_i}\rangle$ and $\langle \phi_{\mathcal{B}_j}|\phi_{\mathcal{A}_i}\rangle$ during the $S$-$D$ wave mixing analysis. These components are associated with the binding energies of the discussed molecular states. Since these molecules have yet to be observed experimentally, we adopt three representative binding energies, namely $-0.5$ MeV, $-6.0$ MeV, and $-12.0$ MeV, to illustrate the magnetic moments and the transition magnetic moments for the $\Xi_{c}^{\prime}\bar D_s^*$ molecular state with $I(J^P)=1/2({3}/{2}^{-})$ and the $\Xi_{c}^{*}\bar D_s^*$ molecule with $I(J^P)=1/2({5}/{2}^{-})$. The corresponding numerical results are presented in Table~\ref{ME2}.
\renewcommand\tabcolsep{0.24cm}
\renewcommand{\arraystretch}{1.50}
\begin{table}[!htbp]
  \caption{The magnetic moments and the transition magnetic moments of the $\Xi_{c}^{\prime}\bar D_s^*$ molecular state with $I(J^P)=1/2({3}/{2}^{-})$ and the $\Xi_{c}^{*}\bar D_s^*$ molecular state with $I(J^P)=1/2({5}/{2}^{-})$ when performing the $S$-$D$ wave mixing analysis. Given that these discussed molecules have not been observed in experiments, we consider three representative binding energies, namely $-0.5~{\rm MeV}$, $-6.0~{\rm MeV}$, and $-12.0~{\rm MeV}$, for the investigated molecules. The corresponding  magnetic moments and transition magnetic moments are listed in the third column.}
  \label{ME2}
\begin{tabular}{c|c|c}
\toprule[1.0pt]
\toprule[1.0pt]
$I_3$&Physical quantities &  Values \\\hline
\multirow{3}{*}{$\frac{1}{2}$}   &$\mu_{\Xi_c^{\prime}\bar D_s^*|{3}/{2}^-\rangle}$  & $-0.414~\mu_N$,\,$-0.416~\mu_N$,\,$-0.416~\mu_N$ \\
                                 &$\mu_{\Xi_c^{*}\bar D_s^*|{5}/{2}^-\rangle}$  & $0.472~\mu_N$,\,$0.472~\mu_N$,\,$0.472~\mu_N$ \\
                                 &$\mu_{\Xi_c^{*}\bar D_s^*|{5}/{2}^-\rangle \to \Xi_c^{\prime}\bar D_s^*|{3}/{2}^-\rangle}$  & $0.154~\mu_N$,\,$0.154~\mu_N$,\,$0.154~\mu_N$ \\\hline
\multirow{3}{*}{$-\frac{1}{2}$}   &$\mu_{\Xi_c^{\prime}\bar D_s^*|{3}/{2}^-\rangle}$  & $-2.272~\mu_N$,\,$-2.264~\mu_N$,\,$-2.261~\mu_N$ \\
                                 &$\mu_{\Xi_c^{*}\bar D_s^*|{5}/{2}^-\rangle}$  & $-2.320~\mu_N$,\,$-2.319~\mu_N$,\,$-2.319~\mu_N$ \\
                                 &$\mu_{\Xi_c^{*}\bar D_s^*|{5}/{2}^-\rangle \to \Xi_c^{\prime}\bar D_s^*|{3}/{2}^-\rangle}$  & $-0.865~\mu_N$,\,$-0.864~\mu_N$,\,$-0.864~\mu_N$ \\
\bottomrule[1.0pt]
\bottomrule[1.0pt]
\end{tabular}
\end{table}

Regarding the obtained magnetic moments and transition magnetic moments of the $\Xi_{c}^{\prime}\bar D_s^*$ molecule with $I(J^P)=1/2({3}/{2}^{-})$ and the $\Xi_{c}^{*}\bar D_s^*$ molecular state with $I(J^P)=1/2({5}/{2}^{-})$ after incorporating the contribution of the $D$-wave channels, two important points should be emphasized:

(i) The $S$-$D$ wave mixing effect does not significantly influence the magnetic moments and the transition magnetic moments of the $\Xi_{c}^{\prime}\bar D_s^*$ molecular state with $I(J^P)=1/2({3}/{2}^{-})$ and the $\Xi_{c}^{*}\bar D_s^*$ molecule with $I(J^P)=1/2({5}/{2}^{-})$. This is due to the fact that the $S$-wave channels predominantly contribute, with probabilities exceeding 99\%, and play a crucial role in the formation of the loosely bound states for both the $\Xi_{c}^{\prime}\bar D_s^*$ state with $I(J^P)=1/2({3}/{2}^{-})$ and the $\Xi_{c}^{*}\bar D_s^*$ state with $I(J^P)=1/2({5}/{2}^{-})$ \cite{Wang:2020bjt}.

(ii) The electromagnetic properties of the $\Xi_{c}^{\prime}\bar D_s^*$ molecule with $I(J^P)=1/2({3}/{2}^{-})$ and the $\Xi_{c}^{*}\bar D_s^*$ molecular state with $I(J^P)=1/2({5}/{2}^{-})$ exhibit minimal dependence on their respective binding energies, as the relevant mixing channel components are not strongly affected by these binding energies \cite{Wang:2020bjt}.

\subsubsection{The coupled channel analysis}

Finally, we delve into the magnetic moments and the transition magnetic moments of the hidden-charm molecular pentaquarks with double strangeness that were previously discussed, using the coupled channel analysis. Specifically, we focus on the $\Xi_{c}^{\prime}\bar D_s^*$ molecular state with $I(J^P)=1/2({3}/{2}^{-})$. For this state, we consider the contribution of the coupled channel effect arising from the $\Xi_{c}^{\prime}\bar D_s^*$ and $\Xi_{c}^{*}\bar D_s^*$ channels \cite{Wang:2020bjt}.

By taking into account the coupled channel effect involving two channels, denoted as $\mathcal{A}$ and $\mathcal{B}$, we can derive the magnetic moment of the molecular state
\begin{eqnarray}
&&\sum_{i,\,j}\mu_{\mathcal{A}_i\to \mathcal{A}_j}\langle \phi_{\mathcal{A}_j}|\phi_{\mathcal{A}_i}\rangle+\sum_{i,\,j}\mu_{\mathcal{B}_i\to \mathcal{B}_j}\langle \phi_{\mathcal{B}_j}|\phi_{\mathcal{B}_i}\rangle\nonumber\\
&&+\sum_{i,\,j} \mu_{\mathcal{B}_i \to \mathcal{A}_j}\langle \phi_{\mathcal{A}_j}|\phi_{\mathcal{B}_i}\rangle+\sum_{i,\,j} \mu_{\mathcal{A}_i \to \mathcal{B}_j}\langle \phi_{\mathcal{B}_j}|\phi_{\mathcal{A}_i}\rangle,
\end{eqnarray}
while the transition magnetic moment between the molecular states can be given by
\begin{eqnarray}
&&\sum_{i,\,j} \mu_{\mathcal{A}_i \to \mathcal{C}_j}\langle \phi_{\mathcal{C}_j}|\phi_{\mathcal{A}_i}\rangle+\sum_{i,\,j} \mu_{\mathcal{A}_i \to \mathcal{D}_j}\langle \phi_{\mathcal{D}_j}|\phi_{\mathcal{A}_i}\rangle\nonumber\\
&&+\sum_{i,\,j} \mu_{\mathcal{B}_i \to \mathcal{C}_j}\langle \phi_{\mathcal{C}_j}|\phi_{\mathcal{B}_i}\rangle+\sum_{i,\,j} \mu_{\mathcal{B}_i \to \mathcal{D}_j}\langle \phi_{\mathcal{D}_j}|\phi_{\mathcal{B}_i}\rangle.
\end{eqnarray}

After performing extensive and intricate calculations, we can determine the magnetic moments and the transition magnetic moments of the hidden-charm molecular pentaquarks with double strangeness by utilizing the coupled channel analysis. The corresponding numerical results are compiled in Table~\ref{ME3}. In order to present a comprehensive picture, we take three representative binding energies, namely $-0.5$ MeV, $-6.0$ MeV, and $-12.0$ MeV, for the discussed molecular states to present the corresponding numerical results.

\renewcommand\tabcolsep{0.45cm}
\renewcommand{\arraystretch}{1.50}
\begin{table}[!htbp]
  \caption{The magnetic moments and the transition magnetic moments of these discussed hidden-charm molecular pentaquarks with double strangeness when performing the coupled channel analysis. Given that these discussed molecules have not yet been observed in experiments, we present the corresponding numerical results for the investigated molecules using three representative binding energies: $-0.5~{\rm MeV}$, $-6.0~{\rm MeV}$, and $-12.0~{\rm MeV}$, respectively, as listed in the second column.}\label{ME3}
\begin{tabular}{c|c}
\toprule[1.0pt]
\toprule[1.0pt]
Physical quantities &  Values \\\hline
$\mu_{\Xi_c^{\prime}\bar D_s^*|{3}/{2}^-\rangle}^{I_3=1/2}$  & $-0.389~\mu_N$,\,$-0.335~\mu_N$,\,$-0.308~\mu_N$ \\
$\mu_{\Xi_c^{\prime}\bar D_s^*|{3}/{2}^-\rangle}^{I_3=-1/2}$  & $-2.305~\mu_N$,\,$-2.354~\mu_N$,\,$-2.370~\mu_N$ \\
$\mu_{\Xi_c^{*}\bar D_s^*|{5}/{2}^-\rangle \to \Xi_c^{\prime}\bar D_s^*|{3}/{2}^-\rangle}^{I_3=1/2}$  & $0.088~\mu_N$,\,$-0.039~\mu_N$,\,$-0.091~\mu_N$ \\
$\mu_{\Xi_c^{*}\bar D_s^*|{5}/{2}^-\rangle \to \Xi_c^{\prime}\bar D_s^*|{3}/{2}^-\rangle}^{I_3=-1/2}$  & $-0.866~\mu_N$,\,$-0.864~\mu_N$,\,$-0.862~\mu_N$ \\
\bottomrule[1.0pt]
\bottomrule[1.0pt]
\end{tabular}
\end{table}

Upon incorporating the contribution of the coupled channel effect, the magnetic moments and the transition magnetic moments of the hidden-charm molecular pentaquarks with double strangeness undergo modifications. One notable alteration is observed in the transition magnetic moment of the $\Xi_c^{*}\bar D_s^*|{5}/{2}^-\rangle \to \Xi_c^{\prime}\bar D_s^*|{3}/{2}^-\rangle$ process with $I_3=1/2$, which can attain a substantial value of $0.245~\mu_N$.

Considering that the thresholds of the $\Xi_{c}^{*}\bar D_s^*$ and $\Omega_{c}^{*}\bar D^*$ channels are in close proximity, the magnetic moment of the $\Xi_{c}^{*}\bar D_s^*$ molecule with $I(J^P)=1/2({5}/{2}^{-})$ may be influenced by the mixing with the $\Omega_{c}^{*}\bar D^*$ channel. In the subsequent analysis, we investigate the magnetic moment of the $\Xi_{c}^{*}\bar D_s^*$ molecule with $I(J^P)=1/2({5}/{2}^{-})$ while taking into account the mixing between the $\Xi_{c}^{*}\bar D_s^*$ and $\Omega_{c}^{*}\bar D^*$ channels. As a crucial piece of input information, it is imperative to examine the mass spectrum of the coupled channel system comprising $\Xi_{c}^{*}\bar D_s^*/\Omega_{c}^{*}\bar D^*$. In order to derive the effective potentials in the coordinate space for the $\Xi_{c}^{*}\bar D_s^*/\Omega_{c}^{*}\bar D^*$ coupled channel system, we employ the one-boson-exchange model in our calculations \cite{Chen:2016qju}. Initially, we express the scattering amplitudes $\mathcal{M}^{h_1h_2\to h_3h_4}$ for the $h_1h_2\to h_3h_4$ scattering processes using the effective Lagrangian approach. The relevant effective Lagrangians, which describe the coupling of the heavy hadrons $\mathcal{B}_{6}^{*}/{\bar D}^{*}$ with the light mesons, are constructed as follows  \cite{Wise:1992hn,Casalbuoni:1992gi,Casalbuoni:1996pg,Yan:1992gz,Bando:1987br,Harada:2003jx,Ding:2008gr,Chen:2017xat}:
\begin{eqnarray}
\mathcal{L}_{\mathcal{B}_{6}^{*}\mathcal{B}_{6}^{*}\sigma} &=&l_S\langle\bar{\mathcal{B}}_{6\mu}^{*}\sigma\mathcal{B}_6^{*\mu}\rangle,\label{n1}\\
\mathcal{L}_{\mathcal{B}_6^{*}\mathcal{B}_6^{*}\mathbb{P}} &=&-i\frac{3g_1}{2f_{\pi}}\varepsilon^{\mu\nu\lambda\kappa}v_{\kappa}\langle\bar{\mathcal{B}}_{6\mu}^{*}\partial_{\nu}\mathbb{P}\mathcal{B}_{6\lambda}^*\rangle,\\
\mathcal{L}_{\mathcal{B}_6^{*}\mathcal{B}_6^{*}\mathbb{V}}&=&\frac{\beta_Sg_V}{\sqrt{2}}\langle\bar{\mathcal{B}}_{6\mu}^*v\cdot \mathbb{V}\mathcal{B}_6^{*\mu}\rangle\nonumber\\
    &&+i\frac{\lambda_Sg_V}{\sqrt{2}}\langle\bar{\mathcal{B}}_{6\mu}^*\left(\partial^{\mu}\mathbb{V}^{\nu}-\partial^{\nu}\mathbb{V}^{\mu}\right)\mathcal{B}_{6\nu}^*\rangle,\\
\mathcal{L}_{{\bar D}^{*}{\bar D}^{*}\sigma} &=&2g_S {\bar D}_{a\mu}^* {\bar D}_a^{*\mu\dag} \sigma,\label{n2}\\
\mathcal{L}_{{\bar D}^{*}{\bar D}^{*}\mathbb{P}}&=&\frac{2ig}{f_{\pi}}v^{\alpha}\varepsilon_{\alpha\mu\nu\lambda}{\bar D}_a^{*\mu\dag}{\bar D}_b^{*\lambda}\partial^{\nu}{\mathbb{P}}_{ab},\\
\mathcal{L}_{{\bar D}^{*}{\bar D}^{*}\mathbb{V}} &=&-\sqrt{2}\beta g_V {\bar D}_{a\mu}^* {\bar D}_b^{*\mu\dag}v\cdot\mathbb{V}_{ab}\nonumber\\
    &&-2\sqrt{2}i\lambda g_V {\bar D}_a^{*\mu\dag}{\bar D}_b^{*\nu}\left(\partial_{\mu}\mathbb{V}_{\nu}-\partial_{\nu}\mathbb{V}_{\mu}\right)_{ab}.
\end{eqnarray}
Here, $v=(1,\bm{0})$ is the four velocity under the nonrelativistic approximation, and the matrices $\mathcal{B}_6^{*}$, ${\mathbb{P}}$, and $\mathbb{V}_{\mu}$ can be written as
\begin{eqnarray*}
\mathcal{B}_6^{*} = \left(\begin{array}{ccc}
         \Sigma_c^{{*}++}                  &\frac{\Sigma_c^{{*}+}}{\sqrt{2}}     &\frac{\Xi_c^{*+}}{\sqrt{2}}\\
         \frac{\Sigma_c^{{*}+}}{\sqrt{2}}      &\Sigma_c^{{*}0}    &\frac{\Xi_c^{*0}}{\sqrt{2}}\\
         \frac{\Xi_c^{*+}}{\sqrt{2}}    &\frac{\Xi_c^{*0}}{\sqrt{2}}      &\Omega_c^{*0}
\end{array}\right),
\end{eqnarray*}
\begin{eqnarray*}
{\mathbb{P}} = {\left(\begin{array}{ccc}
       \frac{\pi^0}{\sqrt{2}}+\frac{\eta}{\sqrt{6}} &\pi^+ &K^+\\
       \pi^-       &-\frac{\pi^0}{\sqrt{2}}+\frac{\eta}{\sqrt{6}} &K^0\\
       K^-         &\bar K^0   &-\sqrt{\frac{2}{3}} \eta     \end{array}\right)},
\end{eqnarray*}
\begin{eqnarray*}
{\mathbb{V}}_{\mu} = {\left(\begin{array}{ccc}
       \frac{\rho^0}{\sqrt{2}}+\frac{\omega}{\sqrt{2}} &\rho^+ &K^{*+}\\
       \rho^-       &-\frac{\rho^0}{\sqrt{2}}+\frac{\omega}{\sqrt{2}} &K^{*0}\\
       K^{*-}         &\bar K^{*0}   & \phi     \end{array}\right)}_{\mu},
\end{eqnarray*}
respectively. In addition, these coupling constants are $l_S=6.20$, $g_S=0.76$, $g_1=0.94$, $g=0.59$, $f_\pi=132~\rm{MeV}$, $\beta_S g_V=10.14$, $\beta g_V=-5.25$, $\lambda_S g_V=19.2~\rm{GeV}^{-1}$, and $\lambda g_V =-3.27~\rm{GeV}^{-1}$ \cite{Chen:2019asm}. And then, the effective potentials in the momentum space $\mathcal{V}^{h_1h_2\to h_3h_4}_E(\bm{q})$ can be related to the scattering amplitudes in terms of the Breit approximation, which can be written as $\mathcal{V}^{h_1h_2\to h_3h_4}_E(\bm{q})=-{\mathcal{M}^{h_1h_2\to h_3h_4}}/{\sqrt{2m_{h_1} 2m_{h_2} 2m_{h_3} 2m_{h_4}}}$. Finally, the effective potentials in the coordinate space $\mathcal{V}^{h_1h_2\to h_3h_4}_E(\bm{r})$ can be obtained by performing the Fourier transform, i.e., $\mathcal{V}^{h_1h_2\to h_3h_4}_E(\bm{r}) =\int \frac{d^3\bm{q}}{(2\pi)^3}e^{i\bm{q}\cdot\bm{r}}\mathcal{V}^{h_1h_2\to h_3h_4}_E(\bm{q})\mathcal{F}^2(q^2,m_E^2)$. In order to account for the finite size of the discussed hadrons, we introduce the monopole-type form factor $\mathcal{F}(q^2,m_E^2) = (\Lambda^2-m_E^2)/(\Lambda^2-q^2)$ in each interaction vertex. This form factor accounts for the nonpointlike nature of the particles involved. Using this standard approach, we deduce the effective potentials in the coordinate space for the $\Xi_{c}^{*}\bar D_s^*/\Omega_{c}^{*}\bar D^*$ coupled channel system, which incorporate the following interactions:
\begin{eqnarray}
\mathcal{V}^{\Xi_{c}^*\bar D_s^*\rightarrow\Xi_{c}^{*}\bar D_s^*}&=&-l_Sg_S\mathcal{A}_{1}Y_\sigma-\frac{g_1 g}{12f_\pi^2}\left[\mathcal{A}_{2}\mathcal{O}_r+\mathcal{A}_{3}\mathcal{P}_r\right]Y_{\eta}\nonumber\\
&&-\frac{\beta \beta_S g_{V}^2}{4}\mathcal{A}_{1}Y_{\phi}\nonumber\\
&&+\frac{\lambda \lambda_S g_V^2}{6}\left[2\mathcal{A}_{2}\mathcal{O}_r-\mathcal{A}_{3}\mathcal{P}_r\right]Y_{\phi},\\
\mathcal{V}^{\Xi_{c}^*\bar D_s^*\rightarrow\Omega_{c}^*\bar D^*}&=&-\frac{g_1 g}{2\sqrt{2}f_\pi^2}\left[\mathcal{A}_{2}\mathcal{O}_r+\mathcal{A}_{3}\mathcal{P}_r\right]Y_{K0}\nonumber\\
&&-\frac{\beta \beta_S g_{V}^2}{2\sqrt{2}}\mathcal{A}_{1}Y_{K^{*}0}\nonumber\\
&&+\frac{\lambda \lambda_S g_V^2}{3\sqrt{2}}\left[2\mathcal{A}_{2}\mathcal{O}_r-\mathcal{A}_{3}\mathcal{P}_r\right]Y_{K^{*}0},\\
\mathcal{V}^{\Omega_{c}^*\bar D^*\rightarrow\Omega_{c}^*\bar D^*}&=&\frac{g_1 g}{6f_\pi^2}\left[\mathcal{A}_{2}\mathcal{O}_r+\mathcal{A}_{3}\mathcal{P}_r\right]Y_{\eta},
\end{eqnarray}
where $\mathcal{O}_r = \frac{1}{r^2}\frac{\partial}{\partial r}r^2\frac{\partial}{\partial r}$, $\mathcal{P}_r = r\frac{\partial}{\partial r}\frac{1}{r}\frac{\partial}{\partial r}$, and the function $Y_i$ is defined as
\begin{eqnarray}
Y_i= \dfrac{e^{-m_ir}-e^{-\Lambda_ir}}{4\pi r}-\dfrac{\Lambda_i^2-m_i^2}{8\pi\Lambda_i}e^{-\Lambda_ir}.
\end{eqnarray}
Here, $m_i=\sqrt{m^2-q_i^2}$, $\Lambda_i=\sqrt{\Lambda^2-q_i^2}$, and $q_0=0.113~{\rm GeV}$. In the above effective potentials, we introduce three operators, i.e.,
\begin{eqnarray*}
\mathcal{A}_{1}&=&\sum_{a,b,m,n}C^{\frac{3}{2},a+b}_{\frac{1}{2}a,1b}C^{\frac{3}{2},m+n}_{\frac{1}{2}m,1n}\chi^{\dagger a}_3\left({\bm\epsilon^{n}_{1}}\cdot{\bm\epsilon^{\dagger b}_{3}}\right)\left({\bm\epsilon_{2}}\cdot{\bm\epsilon^{\dagger}_{4}}\right)\chi^m_1,\nonumber\\
\mathcal{A}_{2}&=&\sum_{a,b,m,n}C^{\frac{3}{2},a+b}_{\frac{1}{2}a,1b}C^{\frac{3}{2},m+n}_{\frac{1}{2}m,1n}\chi^{\dagger a}_3\left({\bm\epsilon^{n}_{1}}\times{\bm\epsilon^{\dagger b}_{3}}\right)\cdot\left({\bm\epsilon_{2}}\times{\bm\epsilon^{\dagger}_{4}}\right)\chi^m_1,\nonumber\\
\mathcal{A}_{3}&=&\sum_{a,b,m,n}C^{\frac{3}{2},a+b}_{\frac{1}{2}a,1b}C^{\frac{3}{2},m+n}_{\frac{1}{2}m,1n}\chi^{\dagger a}_3S({\bm\epsilon^{n}_{1}}\times{\bm\epsilon^{\dagger b}_{3}},{\bm\epsilon_{2}}\times{\bm\epsilon^{\dagger}_{4}},\hat{\bm r})\chi^m_1,
\end{eqnarray*}
where $S({\bm x},{\bm y},\hat{\bm r})=3\left(\hat{\bm r} \cdot {\bm x}\right)\left(\hat{\bm r} \cdot {\bm y}\right)-{\bm x} \cdot {\bm y}$ is the tensor-force operator. In the concrete calculations, these operators should be sandwiched by the spin-orbital wave functions of the initial state and the final state, and the corresponding operator matrix elements with $J=5/2$ are
\begin{eqnarray*}
\mathcal{A}_{1}&=&{\rm diag}(1,1,1,1),\nonumber\\
\mathcal{A}_{2}&=&{\rm diag}\left(-1,\frac{5}{3},\frac{2}{3},-1\right),\nonumber\\
\mathcal{A}_{3}&=&\left(\begin{array}{cccc} 0 & \frac{2}{\sqrt{15}}& \frac{\sqrt{7}}{5\sqrt{3}}& -\frac{2\sqrt{14}}{5}\\ \frac{2}{\sqrt{15}} & 0& \frac{\sqrt{7}}{3\sqrt{5}} & -\frac{4\sqrt{2}}{\sqrt{105}} \\ \frac{\sqrt{7}}{5\sqrt{3}} & \frac{\sqrt{7}}{3\sqrt{5}}& -\frac{16}{21}& -\frac{\sqrt{2}}{7\sqrt{3}} \\-\frac{2\sqrt{14}}{5}&-\frac{4\sqrt{2}}{\sqrt{105}} &-\frac{\sqrt{2}}{7\sqrt{3}}&-\frac{4}{7}\end{array}\right).
\end{eqnarray*}
Based on the obtained effective potentials in the coordinate space for the $\Xi_{c}^{*}\bar D_s^*/\Omega_{c}^{*}\bar D^*$ coupled channel system, we can discuss the bound state properties for the $\Xi_{c}^{*}\bar D_s^*/\Omega_{c}^{*}\bar D^*$ coupled channel system with $I(J^P)=1/2({5}/{2}^{-})$ by solving the coupled channel Schr$\ddot{\rm o}$dinger equation. Table~\ref{reply1} presents the solutions for the bound states in the $\Xi_{c}^{*}\bar D_s^*/\Omega_{c}^{*}\bar D^*$ coupled channel system with $I(J^P)=1/2({5}/{2}^{-})$. Notably, this calculation also provides the probabilities associated with these involved channels, which serve as the crucial input information for the discussion of the magnetic moment of the $\Xi_{c}^{*}\bar D_s^*/\Omega_{c}^{*}\bar D^*$ coupled channel system with $I(J^P)=1/2({5}/{2}^{-})$.

\renewcommand\tabcolsep{0.40cm}
\renewcommand{\arraystretch}{1.50}
\begin{table}[!htbp]
\caption{Bound state solutions for the $\Xi_{c}^{*}\bar D_s^*/\Omega_{c}^{*}\bar D^*$ coupled channel system with $I(J^P)=1/2({5}/{2}^{-})$.}\label{reply1}
\centering
\begin{tabular}{ccccc}\toprule[1pt]\toprule[1pt]
$\Lambda~({\rm GeV})$ &$E~({\rm MeV})$  &$r_{\rm RMS}~({\rm fm})$ &P($\Xi_{c}^{*}\bar D_s^*/\Omega_{c}^{*}\bar D^*$)\\
\hline
1.544&$-0.536$ &3.746&\textbf{97.129}/2.871\\
1.575&$-6.086$ &1.154&\textbf{88.923}/11.077\\
1.593&$-12.188$ &0.813&\textbf{84.560}/15.440\\
\bottomrule[1pt]\bottomrule[1pt]
\end{tabular}
\end{table}

Additionally, we obtain the magnetic moments $\mu_{\Omega_c^{0}}=-1.018~\mu_N$, $\mu_{\bar D^{0}}=1.489~\mu_N$, and $\mu_{D^{-}}=-1.303~\mu_N$ using the constituent quark model. In Table~\ref{reply2}, we compare the calculated magnetic moments of the pure $\Xi_{c}^{*}\bar D_s^*$ molecule with  $I(J^P)=1/2({5}/{2}^{-})$ and the mixed $\Xi_{c}^{*}\bar D_s^*/\Omega_{c}^{*}\bar D^*$ molecule with $I(J^P)=1/2({5}/{2}^{-})$. The obtained results demonstrate that the influence of the $\Omega_{c}^{*}\bar D^*$ channel on the magnetic moment of the $\Xi_{c}^{*}\bar D_s^*$ molecule with $I(J^P)=1/2({5}/{2}^{-})$ is negligible. This can be attributed to the dominant contribution of the $\Xi_{c}^{*}\bar D_s^*$ channel, with a probability exceeding 80\%, and the magnetic moment of the $\Xi_{c}^{*}\bar D_s^*$ state with $I(J^P)=1/2({5}/{2}^{-})$ being remarkably close to that of the $\Omega_{c}^{*}\bar D^*$ state with $I(J^P)=1/2({5}/{2}^{-})$.

\renewcommand\tabcolsep{0.75cm}
\renewcommand{\arraystretch}{1.50}
\begin{table}[!htbp]
  \caption{The magnetic moments of the pure $\Xi_{c}^{*}\bar D_s^*$ molecule with $I(J^P)=1/2({5}/{2}^{-})$ and the mixed $\Xi_{c}^{*}\bar D_s^*/\Omega_{c}^{*}\bar D^*$ molecule with $I(J^P)=1/2({5}/{2}^{-})$.}
  \label{reply2}
\centering
\begin{tabular}{c|c|c}
\toprule[1.0pt]
\toprule[1.0pt]
$I_3(J^P)$&$\mu_{\Xi_c^{*}\bar D_s^*}$ &  $\mu_{\Xi_c^{*}\bar D_s^*/\Omega_{c}^{*}\bar D^*}$ \\\hline
$1/2({5}/{2}^-)$ & $0.472~\mu_N$ & $0.472~\mu_N$ \\
$-1/2({5}/{2}^-)$ & $-2.321~\mu_N$& $-2.321~\mu_N$\\
\bottomrule[1.0pt]
\bottomrule[1.0pt]
\end{tabular}
\end{table}

\subsection{The radiative decay behavior of the  $\Xi_c^{(\prime,*)} \bar{D}_s^{*}$ molecules}

The radiative decay process provides an ideal platform for studying the electromagnetic properties of hadrons experimentally. In the following analysis, we estimate the radiative decay behavior between the $\Xi_{c}^{\prime}\bar D_s^*$ molecular state with $I(J^P)=1/2({3}/{2}^{-})$ and the $\Xi_{c}^{*}\bar D_s^*$ molecule with $I(J^P)=1/2({5}/{2}^{-})$. In this process, denoted as $H \to H^{\prime}\gamma$, the decay width $\Gamma_{H \to H^{\prime}\gamma}$ can be directly related to the corresponding transition magnetic moment $\mu_{H \to H^{\prime}}$ \cite{Dey:1994qi,Simonis:2018rld,Gandhi:2019bju,Hazra:2021lpa,Li:2021ryu,Zhou:2022gra,Wang:2022tib,Rahmani:2020pol,Menapara:2022ksj,Menapara:2021dzi,Gandhi:2018lez,Majethiya:2011ry,Majethiya:2009vx,Shah:2016nxi,Ghalenovi:2018fxh}. The general relation can be expressed as
\begin{eqnarray}
 \Gamma_{H \to H^{\prime}\gamma}=\frac{E_{\gamma}^{3}}{M_{P}^{2}} \frac{\alpha_{\rm {EM}}}{2J_{H}+1}\frac{\sum\limits_{J_{H^{\prime}z},J_{Hz}}\left(\begin{array}{ccc} J_{H^{\prime}}&1&J_{H}\\-J_{H^{\prime}z}&0&J_{Hz}\end{array}\right)^2}{\left(\begin{array}{ccc} J_{H^{\prime}}&1&J_{H}\\-J_{z}&0&J_{z}\end{array}\right)^2}\frac{\left|\mu_{H \to H^{\prime}}\right|^2}{\mu_N^2},\nonumber\\
\end{eqnarray}
where $J_z={\rm Min}\{J_H,\,J_{H^{\prime}}\}$, the notation $\left(\begin{array}{ccc} a&b&c\\d&e&f\end{array}\right)$ stands for the 3$j$ coefficient, $\alpha_{\rm {EM}}$ is the electromagnetic fine structure constant with $\alpha_{\rm {EM}} \approx {1}/{137}$, the proton mass $M_P$ is taken to be $0.938\,\mathrm{GeV}$ \cite{Workman:2022ynf}, and $E_{\gamma}$ is the photon momentum, which is defined by
\begin{equation*}
E_{\gamma}=\frac{M_{H}^2-M_{H^{\prime}}^2}{2M_{H}}.
\end{equation*}
The derivation of the formula for the radiative decay width associated with the corresponding transition magnetic moment is provided in Appendix \ref{app01}. In this study, we specifically investigate the $H({5}/{2}^-) \to H^{\prime} ({3}/{2}^-)\gamma$ process, and the radiative decay width $\Gamma_{H({5}/{2}^-) \to H^{\prime} ({3}/{2}^-)\gamma}$ can be simplified as follows:
\begin{eqnarray}
 \Gamma_{H({5}/{2}^-) \to H^{\prime} ({3}/{2}^-)\gamma}=\alpha_{\rm {EM}}\frac{E_{\gamma}^{3}}{M_{P}^{2}} \frac{5}{6}\frac{\left|\mu_{H \to H^{\prime}}\right|^2}{\mu_N^2}.
\end{eqnarray}

In the previous subsection, we did not consider the contribution of the spatial wave functions of the initial and final states when discussing the transition magnetic moments between the $\Xi_{c}^{\prime}\bar D_s^*$ molecular state with $I(J^P)=1/2({3}/{2}^{-})$ and the $\Xi_{c}^{*}\bar D_s^*$ molecular state with $I(J^P)=1/2({5}/{2}^{-})$. If the momentum of the emitted photon is extremely small, the spatial wave function of the emitted photon, denoted as $e^{-i{\bf q} \cdot{\bf r}_j}$, is approximately equal to 1. As a result, the spatial wave functions of the initial and final states do not significantly affect the final results of the transition magnetic moment and the radiative decay width when the momentum of the emitted photon is particularly small and the overlap of the spatial wave functions of the initial and final hadrons is approximately equal to 1. This approximation has been widely used to discuss the transition magnetic moments and the radiative decay widths of transitions between baryons or mesons, as demonstrated in previous references \cite{Majethiya:2009vx,Majethiya:2011ry,Shah:2016nxi,Gandhi:2018lez,Simonis:2018rld,Ghalenovi:2018fxh,Gandhi:2019bju,Rahmani:2020pol,Hazra:2021lpa,Menapara:2021dzi,Menapara:2022ksj}.
For the $\Xi_c^{*}\bar D_s^*|{5}/{2}^-\rangle \to \Xi_c^{\prime}\bar D_s^*|{3}/{2}^-\rangle\gamma$ process, the momentum of the emitted photon is approximately $60~{\rm MeV}$. Therefore, in this subsection, we will discuss the contribution of the spatial wave functions of the initial and final states to the transition magnetic moments and the radiative decay widths of the $\Xi_c^{*}\bar D_s^*|{5}/{2}^-\rangle \to \Xi_c^{\prime}\bar D_s^*|{3}/{2}^-\rangle\gamma$ process. We will introduce how to account for the contribution of the spatial wave functions of the initial and final states in the calculation of the transition magnetic moment and the radiative decay width.

To accurately assess the impact of the spatial wave functions of the initial and final states on the transition magnetic moment and the radiative decay width, it is necessary to incorporate the spatial wave function of the emitted photon, denoted as $e^{-i{\bf q}\cdot{\bf r}_j}$, into the helicity transition amplitude $\mathcal{A}_{J_{fz},J_{iz}}^{M}$ associated with the magnetic operator \cite{Deng:2016stx,Deng:2015bva,Deng:2016ktl}. Thus, the expression for the helicity transition amplitude becomes
\begin{eqnarray}
\mathcal{A}_{J_{fz},J_{iz}}^M=i\sqrt{\frac{E_{\gamma}}{2}}\langle \psi_{f}|\sum_{j}\frac{e_j}{2M_j}\hat{\bm{\sigma}}_{j}\cdot(\bm{\epsilon}\times\hat{\bf q})e^{-i {\bf q}\cdot{\bf r}_j}|\psi_{i}\rangle.
\end{eqnarray}
Hence, the primary objective is to evaluate the matrix element $\langle \psi_{f}|\sum_{j}\frac{e_j}{2M_j}\hat{\sigma}_{zj}e^{-i {\bf q}\cdot{\bf r}_j}|\psi_{i}\rangle$ when considering the contribution of the spatial wave functions of the initial and final states to the transition magnetic moment and the radiative decay width. To simplify the analysis, we just focus on the impact of these spatial wave functions of the emitted photon, the initial hadron molecule, and the final hadron molecule in the context of the general decay process involving the two-body system in the following. For such radiative decay process, we have
\begin{eqnarray}
{\hat{\bm \mu}}={\hat{\bm \mu}_1}+{\hat{\bm \mu}_2}=\mu_1{\hat
{\bf S}}_1+\mu_2{\hat{\bf S}}_2,
\end{eqnarray}
where ${\hat{\bf S}}_1$ and ${\hat {\bf S}}_2$ spin-operators normalized in such a way that $\mu_1$ and $\mu_2$ give the magnetic momenta of particles 1 or 2 (or their transition magnetic momenta) when evaluating over the first or second particle (or their transitions). In this context, the matrix element $\left\langle {\hat{\bm  \mu}}(\bf q)\right\rangle$ can be expressed as
\begin{eqnarray}
\left\langle {\hat{\bm \mu}}(\bf q)\right\rangle&=&\mu_1\langle {\hat{\bf S}}_1 \rangle \int d^3 {\bf r} \phi_f^*({\bf r})\phi_i({\bf r}) e^{i\frac{M_2}{M_1+M_2}{\bf q}\cdot{\bf r}}\nonumber\\
&&+\mu_2 \langle {\hat{\bf S}}_2 \rangle \int d^3 {\bf r} \phi_f^*({\bf r})\phi_i({\bf r}) e^{-i\frac{M_1}{M_1+M_2}{\bf q}\cdot{\bf r}}.
\end{eqnarray}
Here, $\phi_i({\bf r})$ and $\phi_f({\bf r})$ represent the spatial wave functions of the initial and final hadron molecules, respectively. When considering the $S$-wave initial and final hadron molecules, the matrix element $\left\langle {\hat{\bm \mu}}(\bf q)\right\rangle$ can be simplified to
\begin{eqnarray}
\left\langle {\hat{\bm \mu}}(\bf q)\right\rangle
&=&\mu_1 \langle {\hat{\bf S}}_1 \rangle \int d {r} u_f^*({ r})u_i({r}) j_0\left(-\frac{M_2}{M_1+M_2} qr\right)\nonumber\\
&&+\mu_2 \langle {\hat{\bf S}}_2 \rangle \int d {r} u_f^*({ r})u_i({ r})j_0\left(\frac{M_1}{M_1+M_2} qr\right).
\end{eqnarray}
Here, $j_0(x)={\rm sin} x/x$ is the spherical Bessel wave function of $l=0$, while $u_i(r)$ and $u_f(r)$ are the reduced wave functions of the initial and final states, which are related by
\begin{eqnarray}
\phi_i({\bf r})=\frac{u_i(r)}{r}Y_{00}(\hat r)~~~~~{\rm and}~~~~~\phi_f({\bf r})=\frac{u_f(r)}{r}Y_{00}(\hat r).
\end{eqnarray}
For instance, in the case of the $\Xi_c^{*}\bar D_s^*|{5}/{2}^-\rangle \to \Xi_c^{\prime}\bar D_s^*|{3}/{2}^-\rangle \gamma$ process with $I_3=1/2$, the magnetic moment operator is
\begin{eqnarray}
{\hat{\bm \mu}}=\mu_1\hat {{\bf S}}_T+\mu_2\hat {{\bf S}},
\end{eqnarray}
where ${\hat {\bf S}}_T$ is the spin-transition operator and ${\hat {\bf S}}$ is the spin-1 operator. Thus, we can obtain
\begin{eqnarray}
&&\left\langle {\hat{\bm \mu}}(\bf q)\right\rangle^{\Xi_c^{*}\bar D_s^*|{5}/{2}^-\rangle \to \Xi_c^{\prime}\bar D_s^*|{3}/{2}^-\rangle \gamma}_{I_3=1/2}\nonumber\\
&&=\sqrt{\frac{3}{5}}\mu_{\Xi^{*+}_c \to \Xi^{\prime +}_c} \int d {r} u_f^*({r})u_i({r}) j_0\left(-\frac{M_2}{M_1+M_2} qr\right).
\end{eqnarray}
In the given expression, the transition magnetic moment $\mu_{\Xi^{*+}_c \to \Xi^{\prime +}_c}$ is determined to be $0.199~\mu_N$. Therefore, we are required to evaluate the factor $\int d {r} u_f^*({r})u_i({r}) j_0\left(-\frac{M_2}{M_1+M_2} qr\right)$. In the actual calculations, we employ the accurate spatial wave function of the molecular state obtained through quantitative solutions of the Schrödinger equation.  When considering the $S$-wave initial and final hadron molecules for the $\Xi_c^{*}\bar D_s^*|{5}/{2}^-\rangle \to \Xi_c^{\prime}\bar D_s^*|{3}/{2}^-\rangle \gamma$ process with $I_3=1/2$, the overlap of the spatial wave functions of the initial molecule, the final molecule, and the emitted photon $\int d {r} u_f^*({r})u_i({r}) j_0\left(-\frac{M_2}{M_1+M_2} qr\right)$ is expected to be approximately 0.929, 0.988, and 0.992 when considering the binding energies of $-0.5$ MeV, $-6.0$ MeV, and $-12.0$ MeV for the initial and final hadron molecules, respectively. In fact, when discussing the transition magnetic moment and the radiative decay width, it is necessary to account for the contribution of the spatial wave functions of the emitted photon, baryons, mesons, and hadronic molecules. In Appendix \ref{app02}, we provide a detailed discussion regarding the contribution of the spatial wave functions of the emitted photon, baryons, mesons, and hadronic molecules when analyzing the transition magnetic moment and the radiative decay width.

Based on the considerations mentioned above, we can calculate the transition magnetic moments and the radiative decay widths between the $\Xi_{c}^{\prime}\bar D_s^*$ molecular state with $I(J^P)=1/2({3}/{2}^{-})$ and the $\Xi_{c}^{*}\bar D_s^*$ molecule with $I(J^P)=1/2({5}/{2}^{-})$ when taking into account the contribution of the spatial wave functions of the initial and final states. However, since the binding energies of the $\Xi_{c}^{\prime}\bar D_s^*$ and $\Xi_{c}^{*}\bar D_s^*$ molecules are not known experimentally, we consider three representative binding energies: $-0.5$ MeV, $-6.0$ MeV, and $-12.0$ MeV for the initial and final molecules. These choices allow us to present numerical results that span a range of binding energies.

In Table~\ref{TabledecaywidthsPcss}, we provide the transition magnetic moments and the radiative decay widths between the $\Xi_{c}^{\prime}\bar D_s^*$ molecular state with $I(J^P)=1/2({3}/{2}^{-})$ and the $\Xi_{c}^{*}\bar D_s^*$ molecule with $I(J^P)=1/2({5}/{2}^{-})$ after incorporating the contribution of the spatial wave functions of the initial and final states. We analyze these results using three different scenarios: the single channel analysis, the $S$-$D$ wave mixing analysis, and the coupled channel analysis. Each scenario provides insights into the transition magnetic moments and the radiative decay widths from a distinct perspective.

\renewcommand\tabcolsep{0.95cm}
\renewcommand{\arraystretch}{1.50}
\begin{table*}[!htbp]
  \caption{The transition magnetic moments and the radiative decay widths between the $\Xi_{c}^{\prime}\bar D_s^*$ molecule with $I(J^P)=1/2({3}/{2}^{-})$ and the $\Xi_{c}^{*}\bar D_s^*$ molecule with $I(J^P)=1/2({5}/{2}^{-})$ after taking into account the contribution of the spatial wave functions of the initial and final states. Here, the transition magnetic moment is in unit of $\mu_N$, and the radiative decay width is in unit of ${\rm keV}$. Since the $\Xi_{c}^{\prime}\bar D_s^*$ and $\Xi_{c}^{*}\bar D_s^*$ molecules have not been observed experimentally, we consider three representative binding energies: $-0.5~{\rm MeV}$, $-6.0~{\rm MeV}$, and $-12.0~{\rm MeV}$ for these molecules in our calculations. These choices allow us to explore the transition magnetic moments and the radiative decay widths over a range of binding energies. By considering different binding energies, we can assess the sensitivity of the results to the specific nature of the molecular states under investigation.}
  \label{TabledecaywidthsPcss}
\begin{tabular}{c|c|c|c}
\toprule[1.0pt]\toprule[1.0pt]
$I_3$& Single channel analysis& $S$-$D$ wave mixing analysis & Coupled channel analysis\\\hline
\multicolumn{4}{c}{Transition magnetic moments}\\\hline
$1/2$&0.143,\,0.150,\,0.151      &0.143,\,0.150,\,0.151       &0.078,\,$-0.041$,\,$-0.092$  \\
$-1/2$&$-0.816$,\,$-0.854$,\,$-0.856$     &$-0.814$,\,$-0.854$,\,$-0.856$  &$-0.814$,\,$-0.855$,\,$-0.854$  \\\hline
\multicolumn{4}{c}{Radiative decay widths}\\\hline
$1/2$&0.041,\,0.045,\,0.046      &0.041,\,0.045,\,0.045       &0.012,\,0.003,\,0.017  \\
$-1/2$&1.333,\,1.460,\,1.469     &1.327,\,1.463,\,1.467       &1.329,\,1.464,\,1.462  \\
\bottomrule[1.0pt]\bottomrule[1.0pt]
\end{tabular}
\end{table*}

As shown in Table~\ref{TabledecaywidthsPcss}, the radiative decay width of the $\Xi_c^{*}\bar D_s^*|{5}/{2}^-\rangle \to \Xi_c^{\prime}\bar D_s^*|{3}/{2}^-\rangle \gamma$ process with $I_3=1/2$ is much smaller than that with $I_3=-1/2$, which is similar to the radiative decay behavior of the $\Xi^{*+}_c \to \Xi^{\prime +}_c \gamma$ and $\Xi^{*0}_c \to \Xi^{\prime 0}_c \gamma$ processes \cite{Simonis:2018rld,Aliev:2009jt}. Furthermore, the transition magnetic moments and the radiative decay widths of these discussed hidden-charm molecular pentaquarks with double strangeness do not change too much with increasing their binding energies \cite{Wang:2020bjt}. In addition, the $D$-wave component hardly affects the transition magnetic moments and the radiative decay widths of the $\Xi_c^{*}\bar D_s^*|{5}/{2}^-\rangle \to \Xi_c^{\prime}\bar D_s^*|{3}/{2}^-\rangle \gamma$ process.

To illustrate the impact of the spatial wave functions of the emitted photon, baryons, mesons, and hadronic molecules on the transition magnetic moment and the radiative decay width, we compare the obtained transition magnetic moment for the $\Xi_c^{*}\bar D_s^*|{5}/{2}^-\rangle \to \Xi_c^{\prime}\bar D_s^*|{3}/{2}^-\rangle$ process with $I_3=1/2$ under three different cases: (I) Neglecting the spatial wave functions of the emitted photon, baryons, mesons, and hadronic molecules, (II) Considering only the spatial wave functions of the emitted photon and hadronic molecules, and (III) Accounting for the spatial wave functions of the emitted photon, baryons, mesons, and hadronic molecules. The comparison results are presented in Table \ref{Compare}. By examining these results, we observe that the spatial wave functions of the emitted photon, baryons, mesons, and hadronic molecules have a minor effect on the transition magnetic moment of the $\Xi_c^{*}\bar D_s^*|{5}/{2}^-\rangle \to \Xi_c^{\prime}\bar D_s^*|{3}/{2}^-\rangle$ process with $I_3=1/2$. This can be attributed to the relatively small momentum of the emitted photon, which is approximately $60~{\rm MeV}$ for this specific process.
\renewcommand\tabcolsep{1.00cm}
\renewcommand{\arraystretch}{1.50}
\begin{table}[!htbp]
  \caption{The obtained transition magnetic moment of the $\Xi_c^{*}\bar D_s^*|{5}/{2}^-\rangle \to \Xi_c^{\prime}\bar D_s^*|{3}/{2}^-\rangle$ process with $I_3=1/2$ when only considering the $S$-wave component by following three cases: (I) Neglecting the spatial wave functions of the emitted photon, baryons, mesons, and hadronic molecules, (II) Considering only the spatial wave functions of the emitted photon and hadronic molecules, and (III) Accounting for the spatial wave functions of the emitted photon, baryons, mesons, and hadronic molecules. Since the discussed molecules are not yet observed in experiments, we present the corresponding numerical results for three representative binding energies: $-0.5~{\rm MeV}$, $-6.0~{\rm MeV}$, and $-12.0~{\rm MeV}$. By comparing these results, we aim to assess the impact of the spatial wave functions of the emitted photon, baryons, mesons, and hadronic molecules on the transition magnetic moment.}
  \label{Compare}
\centering
\begin{tabular}{c|c}
\toprule[1.0pt]
\toprule[1.0pt]
Cases&$\mu_{\Xi_c^{*}\bar D_s^*|{5}/{2}^-\rangle \to \Xi_c^{\prime}\bar D_s^*|{3}/{2}^-\rangle}^{I_3=1/2}$ \\\hline
(I)&$0.154~\mu_N$\\
(II)&$0.143~\mu_N$, $0.152~\mu_N$, $0.153~\mu_N$\\
(III)&$0.143~\mu_N$, $0.150~\mu_N$, $0.151~\mu_N$\\
\bottomrule[1.0pt]
\bottomrule[1.0pt]
\end{tabular}
\end{table}

\section{The electromagnetic properties of the  $\Omega_{c}^{(*)}\bar D_s^{*}$ molecules}\label{sec3}

In our previous study \cite{Wang:2021hql}, we have predicted the existence of two molecular states: the $\Omega_{c}\bar D_s^*$ molecule with $I(J^P)=0({3}/{2}^{-})$ and the $\Omega_{c}^{*}\bar D_s^*$ molecule with $I(J^P)=0({5}/{2}^{-})$. To gain insights into the inner structures of these molecular states, it is crucial to investigate their electromagnetic properties.

Within the framework of the constituent quark model, the procedure for calculating the magnetic moments and the transition magnetic moments of the $\Omega_{c}^{(*)}\bar D_s^{*}$ molecular states is analogous to that of the $\Xi_c^{(\prime,*)} \bar D_s^*$ molecular states. The flavor wave functions of the $\Omega_{c}^{(*)0}$ baryons can be expressed as $ssc$, where $s$ represents the strange quark and $c$ denotes the charm quark. The corresponding spin wave functions $|S,S_3\rangle$ can be written as
\begin{eqnarray*}
\Omega_{c}&:&\left\{
  \begin{array}{l}
    \left|\dfrac{1}{2},\dfrac{1}{2}\right\rangle=\dfrac{1}{\sqrt{6}}\left(2\uparrow\uparrow\downarrow-\downarrow\uparrow\uparrow-\uparrow\downarrow\uparrow\right)\\
    \left|\dfrac{1}{2},-\dfrac{1}{2}\right\rangle=\dfrac{1}{\sqrt{6}}\left(\downarrow\uparrow\downarrow+\uparrow\downarrow\downarrow-2\downarrow\downarrow\uparrow\right)
  \end{array}
\right.,\\
\Omega_{c}^{*}&:&\left\{
  \begin{array}{l}
    \left|\dfrac{3}{2},\dfrac{3}{2}\right\rangle=\uparrow\uparrow\uparrow\\
    \left|\dfrac{3}{2},\dfrac{1}{2}\right\rangle=\dfrac{1}{\sqrt{3}}\left(\downarrow\uparrow\uparrow+\uparrow\downarrow\uparrow+\uparrow\uparrow\downarrow\right)\\
     \left|\dfrac{3}{2},-\dfrac{1}{2}\right\rangle=\dfrac{1}{\sqrt{3}}\left(\downarrow\downarrow\uparrow+\uparrow\downarrow\downarrow+\downarrow\uparrow\downarrow\right)\\
    \left|\dfrac{3}{2},-\dfrac{3}{2}\right\rangle=\downarrow\downarrow\downarrow
  \end{array}
\right..
\end{eqnarray*}

Table~\ref{MT2} presents the expressions and numerical results of the magnetic moments and the transition magnetic moment of the $\Omega_c^{(*)}$ baryons. We also compare these obtained numerical results with other theoretical predictions \cite{Gandhi:2018lez,Patel:2007gx,Simonis:2018rld,Sharma:2010vv,Majethiya:2009vx}, and find that our obtained values are consistent with those from other theoretical studies \cite{Gandhi:2018lez,Patel:2007gx,Simonis:2018rld,Sharma:2010vv,Majethiya:2009vx}. Notably, the magnetic moments of the $\Omega_c^{0}$ and $\Omega_c^{*0}$ are found to be very close to each other, which is similar to the case of the magnetic moments of the $\Xi_c^{\prime0}$ and $\Xi_c^{*0}$.
\renewcommand\tabcolsep{0.16cm}
\renewcommand{\arraystretch}{1.50}
\begin{table}[!htbp]
  \caption{The magnetic moments and the transition magnetic moment of the $\Omega_c^{(*)}$ baryons. Here, the magnetic moment and the transition magnetic moment are in units of $\mu_N$, and the square brackets in the second column represent the expressions of their magnetic moments and transition magnetic moment.}
  \label{MT2}
\begin{tabular}{c|l|l}
\toprule[1.0pt]
\toprule[1.0pt]
Quantities &  \multicolumn{1}{c|}{Our results}  &  \multicolumn{1}{c}{Other results} \\\hline
$\mu_{\Omega^{0}_c}$                   & $-1.051\,\left[\frac{4}{3}\mu_s-\frac{1}{3}\mu_c\right]$        & $-1.127$ \cite{Gandhi:2018lez},\,$-0.960$ \cite{Patel:2007gx} \\
$\mu_{\Omega^{*0}_c}$                   & $-1.018\,\left[2\mu_s+\mu_c\right]$                            & $-1.127$ \cite{Gandhi:2018lez},\,$-0.936$ \cite{Simonis:2018rld}\\
$\mu_{\Omega^{*0}_c \to \Omega^{0}_c}$    & $-1.006\,\left[\frac{2\sqrt{2}}{3}(\mu_s-\mu_c)\right]$      &$-0.960$ \cite{Sharma:2010vv},\,$-1.128$ \cite{Majethiya:2009vx}\\
\bottomrule[1.0pt]
\bottomrule[1.0pt]
\end{tabular}
\end{table}

We proceed to investigate the magnetic moments, the transition magnetic moment, and the radiative decay width of the $\Omega_c\bar D_s^*$ molecular state with $I(J^P)=0({3}/{2}^{-})$ and the $\Omega_c^{*}\bar D_s^*$ molecule with $I(J^P)=0({5}/{2}^{-})$. In this case, the flavor wave functions $|I,I_3\rangle$ can be written as $|0,0\rangle=|\Omega_{c}^{(*)0}{D}_s^{*-}\rangle$, where $I$ and $I_3$ represent the isospins and the isospin third components of the $\Omega_{c}^{(*)}\bar{D}_s^{*}$ systems, respectively. The spin wave functions $|S,S_3\rangle$ can be constructed by coupling the spin wave functions of the constituent hadrons, as follows
\begin{eqnarray*}
\Omega_{c}\bar{D}_s^{*}:\,|S,S_3\rangle&=&\sum_{S_{\Omega_{c}},S_{\bar{D}_s^{*}}}C^{SS_3}_{\frac{1}{2}S_{\Omega_{c}},1S_{\bar{D}_s^{*}}}\left|\frac{1}{2},S_{\Omega_{c}}\right\rangle\left|1,S_{\bar{D}_s^{*}}\right\rangle,\\
\Omega_{c}^{*}\bar{D}_s^{*}:\,|S,S_3\rangle&=&\sum_{S_{\Omega_{c}^{*}},S_{\bar{D}_s^{*}}}C^{SS_3}_{\frac{3}{2}S_{\Omega_{c}^{*}},1S_{\bar{D}_s^{*}}}\left|\frac{3}{2},S_{\Omega_{c}^{*}}\right\rangle\left|1,S_{\bar{D}_s^{*}}\right\rangle.
\end{eqnarray*}
Here, $S$ and $S_3$ represent the spins and the spin third components of the $\Omega_{c}^{(*)}\bar{D}_s^{*}$ systems, respectively. The subscripts $S_{\Omega_{c}}$, $S_{\Omega_{c}^{*}}$, and $S_{\bar{D}_s^{*}}$ indicate the spin third components of the $\Omega_{c}$, $\Omega_{c}^{*}$, and $\bar{D}_s^{*}$, respectively.

Similar to the case of the hidden-charm molecular pentaquarks with double strangeness, we investigate the electromagnetic properties of the $\Omega_c\bar D_s^*$ molecular state with $I(J^P)=0({3}/{2}^{-})$ and the $\Omega_c^{*}\bar D_s^*$ molecular state with $I(J^P)=0({5}/{2}^{-})$. In our study, we consider three different scenarios: the single channel analysis, the $S$-$D$ wave mixing analysis, and the coupled channel analysis.

When incorporating the $S$-$D$ wave mixing effect, we account for the allowed $S$-wave and $D$-wave channels $|^{2S+1} L_{J}\rangle$ for the $\Omega_c\bar D_s^*$ molecular state with $I(J^P)=0({3}/{2}^{-})$ and the $\Omega_c^{*}\bar D_s^*$ molecular state with $I(J^P)=0({5}/{2}^{-})$ \cite{Wang:2021hql}
\begin{eqnarray*}
&&\Omega_c\bar D_s^{*}|{3}/{2}^-\rangle:~~|^{4} S_{3/2}\rangle,\,|^{2} D_{3/2}\rangle,\,|^{4} D_{3/2}\rangle,\nonumber\\
&&\Omega_c^{*}\bar D_s^{*}|{5}/{2}^-\rangle:~~|^{6} S_{5/2}\rangle,\,|^{2} D_{5/2}\rangle,\,|^{4} D_{5/2}\rangle,\,|^{6} D_{5/2}\rangle.
\end{eqnarray*}
Furthermore, it is also important to consider the contribution of the coupled channel effect for the $\Omega_c\bar D_s^*$ molecular state with $I(J^P)=0({3}/{2}^{-})$ \cite{Wang:2021hql}.

Table~\ref{ME4} provides the numerical results of the electromagnetic properties of the $\Omega_c\bar D_s^*$ molecular state with $I(J^P)=0({3}/{2}^{-})$ and the $\Omega_c^{*}\bar D_s^*$ molecular state with $I(J^P)=0({5}/{2}^{-})$ obtained through the single channel analysis, the $S$-$D$ wave mixing analysis, and the coupled channel analysis. The calculations are performed considering three representative binding energies: $-0.5$ MeV,  $-6.0$ MeV, and $-12.0$ MeV for the discussed molecular states. In the analysis, we also account for the contribution of the spatial wave functions of the initial and final states to determine the transition magnetic moment and the radiative decay width of the $\Omega_c^*\bar D_s^*|{5}/{2}^-\rangle \to \Omega_c\bar D_s^*|{3}/{2}^-\rangle \gamma$ process.
\renewcommand\tabcolsep{0.30cm}
\renewcommand{\arraystretch}{1.50}
\begin{table*}[!htbp]
  \caption{The magnetic moments, the transition magnetic moment, and the radiative decay width of the $\Omega_c\bar D_s^*$ molecular state with $I(J^P)=0({3}/{2}^{-})$ and the $\Omega_c^{*}\bar D_s^*$ molecular state with $I(J^P)=0({5}/{2}^{-})$ by performing the single channel, $S$-$D$ wave mixing, and coupled channel analysis, respectively. For the transition magnetic moment and the radiative decay width of the $\Omega_c^*\bar D_s^*|{5}/{2}^-\rangle \to \Omega_c\bar D_s^*|{3}/{2}^-\rangle \gamma$ process, we consider the contribution of the spatial wave functions of the initial and final states. Given the absence of experimental evidence for the discussed molecules, we consider three representative binding energies, namely $-0.5~{\rm MeV}$, $-6.0~{\rm MeV}$, and $-12.0~{\rm MeV}$, for the analyzed molecular states. These binding energies are utilized to present the corresponding numerical results in order to provide insights into the electromagnetic properties of the $\Omega_c\bar D_s^*$ molecular state with $I(J^P)=0({3}/{2}^{-})$ and the $\Omega_c^{*}\bar D_s^*$ molecular state with $I(J^P)=0({5}/{2}^{-})$.}
  \label{ME4}
\begin{tabular}{c|c|c|c}
\toprule[1.0pt]
\toprule[1.0pt]
Physical quantities &  Single channel analysis& $S$-$D$ wave mixing analysis & Coupled channel analysis\\\hline
$\mu_{\Omega_c\bar D_s^*|{3}/{2}^-\rangle}$  & $-2.118~\mu_N$ & $-2.117~\mu_N$,\,$-2.117~\mu_N$,\,$-2.117~\mu_N$& $-2.152~\mu_N$,\,$-2.190~\mu_N$,\,$-2.199~\mu_N$\\
$\mu_{\Omega_c^*\bar D_s^*|{5}/{2}^-\rangle}$  & $-2.085~\mu_N$ & $-2.084~\mu_N$,\,$-2.084~\mu_N$,\,$-2.084~\mu_N$&/\\
$\mu_{\Omega_c^*\bar D_s^*|{5}/{2}^-\rangle \to \Omega_c\bar D_s^*|{3}/{2}^-\rangle}$  & $-0.739~\mu_N$,\,$-0.768~\mu_N$,\,$-0.770~\mu_N$& $-0.734~\mu_N$,\,$-0.768~\mu_N$,\,$-0.769~\mu_N$&$-0.743~\mu_N$,\,$-0.787~\mu_N$,\,$-0.789~\mu_N$\\
$\Gamma_{\Omega_c^*\bar D_s^*|{5}/{2}^-\rangle \to \Omega_c\bar D_s^*|{3}/{2}^-\rangle \gamma}$  & $1.305\,{\rm keV}$,\,$1.410\,{\rm keV}$,\,$1.415\,{\rm keV}$ & $1.288\,{\rm keV}$,\,$1.408\,{\rm keV}$,\,$1.412\,{\rm keV}$ & $1.320\,{\rm keV}$,\,$1.479\,{\rm keV}$,\,$1.489\,{\rm keV}$\\
\bottomrule[1.0pt]
\bottomrule[1.0pt]
\end{tabular}
\end{table*}

From Table~\ref{ME4}, we can find several interesting results:
\begin{enumerate}
  \item The magnetic moments of the $\Omega_c^{0}$ and $\Omega_c^{*0}$ baryons are found to be very close to each other, which indicates that the magnetic moments of the $\Omega_c\bar D_s^*$ molecular state with $I(J^P)=0({3}/{2}^{-})$ and the $\Omega_c^{*}\bar D_s^*$ molecular state with $I(J^P)=0({5}/{2}^{-})$ are nearly identical. Moreover, the radiative decay width of the $\Omega_c^*\bar D_s^*|{5}/{2}^-\rangle \to \Omega_c\bar D_s^*|{3}/{2}^-\rangle \gamma$ process is estimated to be approximately 1.00 keV.
  \item The $S$-$D$ wave mixing effect has a negligible impact on the magnetic moments, the transition magnetic moment, and the radiative decay width of the $\Omega_c\bar D_s^*$ molecular state with $I(J^P)=0({3}/{2}^{-})$ and the $\Omega_c^{*}\bar D_s^*$ molecular state with $I(J^P)=0({5}/{2}^{-})$. After considering the contribution of the $D$-wave channels, the changes in their magnetic moments and transition magnetic moment are less than $0.005~\mu_N$. The main reason is that the $S$-$D$ wave mixing effect can be ignored for the formation of the $\Omega_c\bar D_s^*$ molecular state with $I(J^P)=0({3}/{2}^{-})$ and the $\Omega_c^{*}\bar D_s^*$ molecular state with $I(J^P)=0({5}/{2}^{-})$ \cite{Wang:2021hql}.
 \item The coupled channel effect has a minor influence on the electromagnetic properties of these discussed hidden-charm molecular pentaquarks with triple strangeness. After considering the coupled channel effect, the changes in their magnetic moments and transition magnetic moment are less than $0.09~\mu_N$.
\end{enumerate}

Furthermore, there are several similarities in the electromagnetic properties of these discussed hidden-charm molecular pentaquarks with double strangeness and triple strangeness. In particular, the numerical results of $\mu_{\Xi_c^{\prime}\bar D_s^*|{3}/{2}^-\rangle}$, $\mu_{\Xi_c^{*}\bar D_s^*|{5}/{2}^-\rangle}$, and $\mu_{\Xi_c^{*}\bar D_s^*|{5}/{2}^-\rangle \to \Xi_c^{\prime}\bar D_s^*|{3}/{2}^-\rangle}$ with $I_3=-1/2$ are close to those  of $\mu_{\Omega_c\bar D_s^*|{3}/{2}^-\rangle}$, $\mu_{\Omega_c^{*}\bar D_s^*|{5}/{2}^-\rangle}$, and $\mu_{\Omega_c^{*}\bar D_s^*|{5}/{2}^-\rangle \to \Omega_c\bar D_s^*|{3}/{2}^-\rangle}$ with $I_3=0$, respectively, since the magnetic moments and the transition magnetic moment of the $\Xi_{c}^{\prime(*)0}$ baryons are similar to those of the $\Omega_{c}^{(*)0}$ baryons.

\section{Summary}\label{sec4}

Since the discovery of the hidden-charm pentaquark structures $P_c(4380)$ and $P_c(4450)$ by the LHCb Collaboration in 2015 \cite{Aaij:2015tga}, the study of the hidden-charm molecular pentaquarks has become a prominent focus in the field of hadron physics \cite{Liu:2013waa,Hosaka:2016pey,Chen:2016qju,Richard:2016eis,Lebed:2016hpi,Olsen:2017bmm,Guo:2017jvc,Liu:2019zoy,Brambilla:2019esw,Chen:2022asf,Meng:2022ozq}. Subsequently, significant progress has been achieved on both the theoretical and experimental fronts in recent years. Researchers have made remarkable advancements in investigating the mass spectrum, decay behavior, and production mechanism of various types of hidden-charm molecular pentaquarks. These investigations have yielded valuable insights to deepen our understanding of the nature of the hidden-charm molecular pentaquarks. However, it is important to note that there is still much more to explore and uncover in this captivating research area.

In previous works \cite{Wang:2020bjt,Wang:2021hql}, the Lanzhou group predicted the existence of the hidden-charm molecular pentaquark candidates with double strangeness and triple strangeness by investigating the $\Xi_c^{(\prime,*)} \bar{D}_s^{(*)}$ and $\Omega_{c}^{(*)}\bar D_s^{(*)}$ interactions, providing their corresponding mass spectra. Motivated by these predictions, our current study aims to investigate the electromagnetic properties of these hidden-charm molecular pentaquark candidates. Specifically, we focus on their magnetic moments, transition magnetic moments, and radiative decay behavior, as these physical quantities offer important insights into their underlying structures. Our analysis takes into account various effects, such as the $S$-$D$ wave mixing effect and the coupled channel effect. It is important to note that this investigation serves as an initial exploration into the electromagnetic properties of the $\Xi_c^{(\prime,*)} \bar{D}_s^{*}$ and $\Omega_{c}^{(*)}\bar D_s^{*}$ molecular states. Further theoretical studies employing different approaches and models are encouraged to delve deeper into this topic. Moreover, the experimental measurements of the electromagnetic properties of these discussed hidden-charm molecular pentaquarks with double strangeness and triple strangeness will pose significant challenges for the future research.

The electromagnetic properties of hadrons play the crucial role in revealing their inner structures, allowing for the distinction of their spin-parity quantum numbers and configurations. While the electromagnetic properties of the hidden-charm molecular pentaquark states with double strangeness and triple strangeness have been studied, it is equally important to investigate the electromagnetic properties of the non-molecular hidden-charm pentaquarks with double strangeness and triple strangeness in the future. Such investigations would contribute to discerning the nature of these hidden-charm pentaquarks more accurately. Currently, our understanding of the electromagnetic properties of the compact hidden-charm pentaquarks with double strangeness and triple strangeness remains limited. Therefore, the further theoretical exploration of their electromagnetic properties is highly encouraged, as it may provide valuable insights for constructing a comprehensive picture of the cluster composed of the hidden-charm pentaquark states with double strangeness and triple strangeness.

\section*{Acknowledgement}

This work is supported by the China National Funds for Distinguished Young Scientists under Grant No. 11825503, National Key Research and Development Program of China under Contract No. 2020YFA0406400, the 111 Project under Grant No. B20063, and the National Natural Science Foundation of China under Grant Nos. 12175091, 11965016, 12047501, and 12247155. F.L.W is also supported by the China Postdoctoral Science Foundation under Grant No. 2022M721440.

\appendix
\section{The derivation of the formula for the radiative decay width related to the transition magnetic moment}\label{app01}

In this appendix, we derive the formula for the radiative decay width related to the corresponding transition magnetic moment. According to Refs. \cite{Deng:2016stx,Deng:2015bva,Deng:2016ktl}, the helicity transition amplitude $\mathcal{A}_{J_{fz},J_{iz}}^{M}$ for the magnetic operator between the initial state $|J_{i},J_{iz}\rangle$ and the final state $|J_{f},J_{fz}\rangle$ can be written as
\begin{eqnarray}
\mathcal{A}_{J_{fz},J_{iz}}^M=i\sqrt{\frac{E_{\gamma}}{2}}\langle J_{f},J_{fz}|\sum_{j}\frac{e_j}{2M_j}\hat{\bm{\sigma}}_{j}\cdot(\bm{\epsilon}\times\hat{\bm{q})}|J_{i},J_{iz}\rangle,
\end{eqnarray}
where $E_{\gamma}$ is the  momentum of the emitted photon. Here, we need to specify that the spatial wave function of the emitted photon $e^{-i{\bm q} \cdot{\bm r}_j}$ was not included in the above expression. For the convenience of calculation, we choose the Pauli spin operator along the $z$ axial, the above expression can be simplified to
\begin{eqnarray}
\mathcal{A}_{J_{fz},J_{iz}}^M=i\sqrt{\frac{E_{\gamma}}{2}}\langle J_{f},J_{fz}|\hat\mu_{z}|J_{i},J_{iz}\rangle
\end{eqnarray}
with $\hat\mu_{z}=\sum_{j}\frac{e_j}{2M_j}\hat{\sigma}_{zj}$. Furthermore, the radiative decay width can be given by \cite{Deng:2016stx,Deng:2015bva,Deng:2016ktl}
\begin{eqnarray}
\Gamma&=&\frac{E_{\gamma}^2}{\pi}\frac{2}{2J_i+1}\sum_{J_{fz},J_{iz}}\left|\mathcal{A}_{J_{fz},J_{iz}}^M\right|^2\nonumber\\
&=&\frac{E_{\gamma}^2}{\pi}\frac{2}{2J_i+1}\sum_{J_{fz},J_{iz}}\frac{E_{\gamma}}{2}\left|\langle J_{f},J_{fz}|\hat\mu_{z}|J_{i},J_{iz}\rangle\right|^2\nonumber\\
&=&\alpha_{\rm {EM}}\frac{E_{\gamma}^3}{M_{P}^{2}}\frac{1}{2J_i+1}\sum\limits_{J_{fz},J_{iz}}\frac{\left|\langle J_{f},J_{fz}|\hat\mu_{z}|J_{i},J_{iz}\rangle\right|^2}{\mu_N^2}.
\end{eqnarray}
In the above expression, we use $\alpha_{\rm {EM}}=e^2/4\pi$ and $\mu_N=e/2M_P$. Thus, the width $\Gamma_{H \to H^{\prime}\gamma}$ of the radiative decay process $H \to H^{\prime}\gamma$ can be expressed as\footnote{Here, we need to specify that the formulas for the radiative decay width related to the transition magnetic moment in Refs. \cite{Wang:2022tib,Zhou:2022gra} can only be used to discuss several special decay processes, such as the $1/2 \to 1/2+\gamma$ process, the $3/2 \to 1/2+\gamma$ process, and so on.}
\begin{equation}
 \Gamma_{H \to H^{\prime}\gamma}=\alpha_{\rm {EM}}\frac{E_{\gamma}^{3}}{M_{P}^{2}} \frac{1}{2J_{H}+1}\sum_{J_{H^{\prime}z},J_{Hz}}\frac{\left|\left\langle{J_{H^{\prime}},J_{H^{\prime}z}|\hat\mu_{z}|J_{H},J_{Hz}}\right\rangle\right|^{2}}{\mu_N^{2}}.
\end{equation}

To explicitly write out the expression for the radiative decay width $\Gamma_{H \to H^{\prime}\gamma}$ related to the corresponding transition magnetic moment $\mu_{H \to H^{\prime}}$, we need to derive the relation between $\left\langle{J_{H^{\prime}},J_{H^{\prime}z}|\hat\mu_{z}|J_{H},J_{Hz}}\right\rangle$ and $\mu_{H \to H^{\prime}}$. According to the Wigner-Eckart theorem \cite{Khersonskii:1988krb}, the expectation value $\left\langle{J_{H^{\prime}},J_{H^{\prime}z}|\hat\mu_{z}|J_{H},J_{Hz}}\right\rangle$ can be written as
\begin{eqnarray}
&&\left\langle{J_{H^{\prime}},J_{H^{\prime}z}|\hat\mu_{z}|J_{H},J_{Hz}}\right\rangle\nonumber\\
&&=(-1)^{J_{H^{\prime}}-J_{H^{\prime}z}}\left(\begin{array}{ccc} J_{H^{\prime}}&1&J_{H}\\-J_{H^{\prime}z}&0&J_{Hz}\end{array}\right)\left\langle{J_{H^{\prime}}||\hat\mu_{z}||J_{H}}\right\rangle,
\end{eqnarray}
where the notation $\left(\begin{array}{ccc} a&b&c\\d&e&f\end{array}\right)$ stands for the 3$j$ coefficient, and the factor $\left\langle{J_{H^{\prime}}||\hat\mu_{z}||J_{H}}\right\rangle$ is called the reduced matrix element, which does not depend on $J_{Hz}$ and $J_{H^{\prime}z}$. From the expression of the transition magnetic moment $\mu_{H \to H^{\prime}}$, we can find
\begin{eqnarray}
\mu_{H \to H^{\prime}}&=&\left\langle{J_{H^{\prime}},J_{z}|\hat\mu_{z}|J_{H},J_{z}}\right\rangle\nonumber\\
&=&(-1)^{J_{H^{\prime}}-J_{z}}\left(\begin{array}{ccc} J_{H^{\prime}}&1&J_{H}\\-J_{z}&0&J_{z}\end{array}\right)\left\langle{J_{H^{\prime}}||\hat\mu_{z}||J_{H}}\right\rangle
\end{eqnarray}
with $J_z={\rm Min}\{J_H,\,J_{H^{\prime}}\}$. Thus, the reduced matrix element $\left\langle{J_{H^{\prime}}||\hat\mu_{z}||J_{H}}\right\rangle$ can be expressed as
\begin{eqnarray}
\left\langle{J_{H^{\prime}}||\hat\mu_{z}||J_{H}}\right\rangle=\frac{\mu_{H \to H^{\prime}}}{(-1)^{J_{H^{\prime}}-J_{z}}\left(\begin{array}{ccc} J_{H^{\prime}}&1&J_{H}\\-J_{z}&0&J_{z}\end{array}\right)}.
\end{eqnarray}
From the above  discussion, we can obtain the relation between $\left\langle{J_{H^{\prime}},J_{H^{\prime}z}|\hat\mu_{z}|J_{H},J_{Hz}}\right\rangle$ and $\mu_{H \to H^{\prime}}$, i.e.,
\begin{eqnarray}
&&\left\langle{J_{H^{\prime}},J_{H^{\prime}z}|\hat\mu_{z}|J_{H},J_{Hz}}\right\rangle\nonumber\\
&&=(-1)^{J_{H^{\prime}}-J_{H^{\prime}z}}\left(\begin{array}{ccc} J_{H^{\prime}}&1&J_{H}\\-J_{H^{\prime}z}&0&J_{Hz}\end{array}\right)\left\langle{J_{H^{\prime}}||\hat\mu_{z}||J_{H}}\right\rangle\nonumber\\
&&=(-1)^{J_{H^{\prime}}-J_{H^{\prime}z}}\left(\begin{array}{ccc} J_{H^{\prime}}&1&J_{H}\\-J_{H^{\prime}z}&0&J_{Hz}\end{array}\right)\frac{\mu_{H \to H^{\prime}}}{(-1)^{J_{H^{\prime}}-J_{z}}\left(\begin{array}{ccc} J_{H^{\prime}}&1&J_{H}\\-J_{z}&0&J_{z}\end{array}\right)}.\nonumber\\
\end{eqnarray}
Considering the relation between $\left\langle{J_{H^{\prime}},J_{H^{\prime}z}|\hat\mu_{z}|J_{H},J_{Hz}}\right\rangle$ and $\mu_{H \to H^{\prime}}$, the radiative decay width $\Gamma_{H \to H^{\prime}\gamma}$ of the $H \to H^{\prime}\gamma$ process can be expressed as
\begin{eqnarray}
 \Gamma_{H \to H^{\prime}\gamma}&=&\alpha_{\rm {EM}}\frac{E_{\gamma}^{3}}{M_{P}^{2}} \frac{1}{2J_{H}+1}\sum_{J_{H^{\prime}z},J_{Hz}}\frac{\left|\left\langle{J_{H^{\prime}},J_{H^{\prime}z}|\hat\mu_{z}|J_{H},J_{Hz}}\right\rangle\right|^{2}}{\mu_N^{2}}\nonumber\\
&=&\frac{E_{\gamma}^{3}}{M_{P}^{2}} \frac{\alpha_{\rm {EM}}}{2J_{H}+1}\frac{\sum\limits_{J_{H^{\prime}z},J_{Hz}}\left(\begin{array}{ccc} J_{H^{\prime}}&1&J_{H}\\-J_{H^{\prime}z}&0&J_{Hz}\end{array}\right)^2}{\left(\begin{array}{ccc} J_{H^{\prime}}&1&J_{H}\\-J_{z}&0&J_{z}\end{array}\right)^2}\frac{\left|\mu_{H \to H^{\prime}}\right|^2}{\mu_N^2}.\nonumber\\
\end{eqnarray}

Based on the preceding discussion, we have obtained the relation between the radiative decay width and the corresponding transition magnetic moment. This connection is established through the utilization of the two 3$j$ coefficients. To further simplify the aforementioned relation, we proceed by evaluating the two 3$j$ coefficients.
In the case of the M1 radiative decay for the $H \rightarrow H^{\prime}\gamma$ process, the total angular momentum quantum numbers of the initial and final hadron states adhere to the condition $J_H=J_{H^{\prime}}$ or $J_H=J_{H^{\prime}}\pm1$. Consequently, by considering these constraints, we are able to determine the specific values of the two 3$j$ coefficients
\begin{eqnarray}
&&\sum\limits_{J_{H^{\prime}z},J_{Hz}}\left(\begin{array}{ccc} J_{H^{\prime}}&1&J_{H}\\-J_{H^{\prime}z}&0&J_{Hz}\end{array}\right)^2=\frac{1}{3},\\
&&\left(\begin{array}{ccc} J_{H^{\prime}}&1&J_{H}\\-J_{z}&0&J_{z}\end{array}\right)^2=\begin{cases}
 \frac{J_H}{(J_H+1)(2J_H+1)}& \text{for}~~J_H=J_{H^{\prime}}, \\
 \frac{1}{J_H(2J_H+1)}& \text{for}~~J_H=J_{H^{\prime}}+1,\\
 \frac{1}{J_{H^{\prime}}(2J_{H^{\prime}}+1)}& \text{for}~~J_H=J_{H^{\prime}}-1.
\end{cases}\nonumber\\
\end{eqnarray}
Based on the aforementioned analysis, we can simplify the relation between the radiative decay width and the corresponding transition magnetic moment to the following expression:
\begin{eqnarray}
 \Gamma_{H \to H^{\prime}\gamma}=\begin{cases}
 \alpha_{\rm {EM}}\frac{E_{\gamma}^{3}}{3M_{P}^{2}} \frac{J_{H}+1}{J_{H}}\frac{\left|\mu_{H \to H^{\prime}}\right|^2}{\mu_N^2}     & \text{for}~~J_H=J_{H^{\prime}}, \\
 \alpha_{\rm {EM}}\frac{E_{\gamma}^{3}}{3M_{P}^{2}} J_{H}\frac{\left|\mu_{H \to H^{\prime}}\right|^2}{\mu_N^2}     & \text{for}~~J_H=J_{H^{\prime}}+1,\\
 \alpha_{\rm {EM}}\frac{E_{\gamma}^{3}}{3M_{P}^{2}} \frac{J_{H^{\prime}}(2J_{H^{\prime}}+1)}{2J_{H}+1}\frac{\left|\mu_{H \to H^{\prime}}\right|^2}{\mu_N^2}     & \text{for}~~J_H=J_{H^{\prime}}-1.
\end{cases}\nonumber\\
\end{eqnarray}
Hence, with the obtained transition magnetic moment, we can now utilize the aforementioned relation to directly examine the radiative decay width of the $H \to H^{\prime}\gamma$ process.

\section{The contribution of the spatial wave functions of the initial and final states for the transition magnetic moment and the radiative decay width}\label{app02}

This appendix provides a comprehensive analysis of the contributions from the spatial wave functions of the emitted photon, baryons, mesons, and hadronic molecules to both the transition magnetic moment and the radiative decay width.

\begin{figure}[htbp]
\centering
\includegraphics[width=7.5cm]{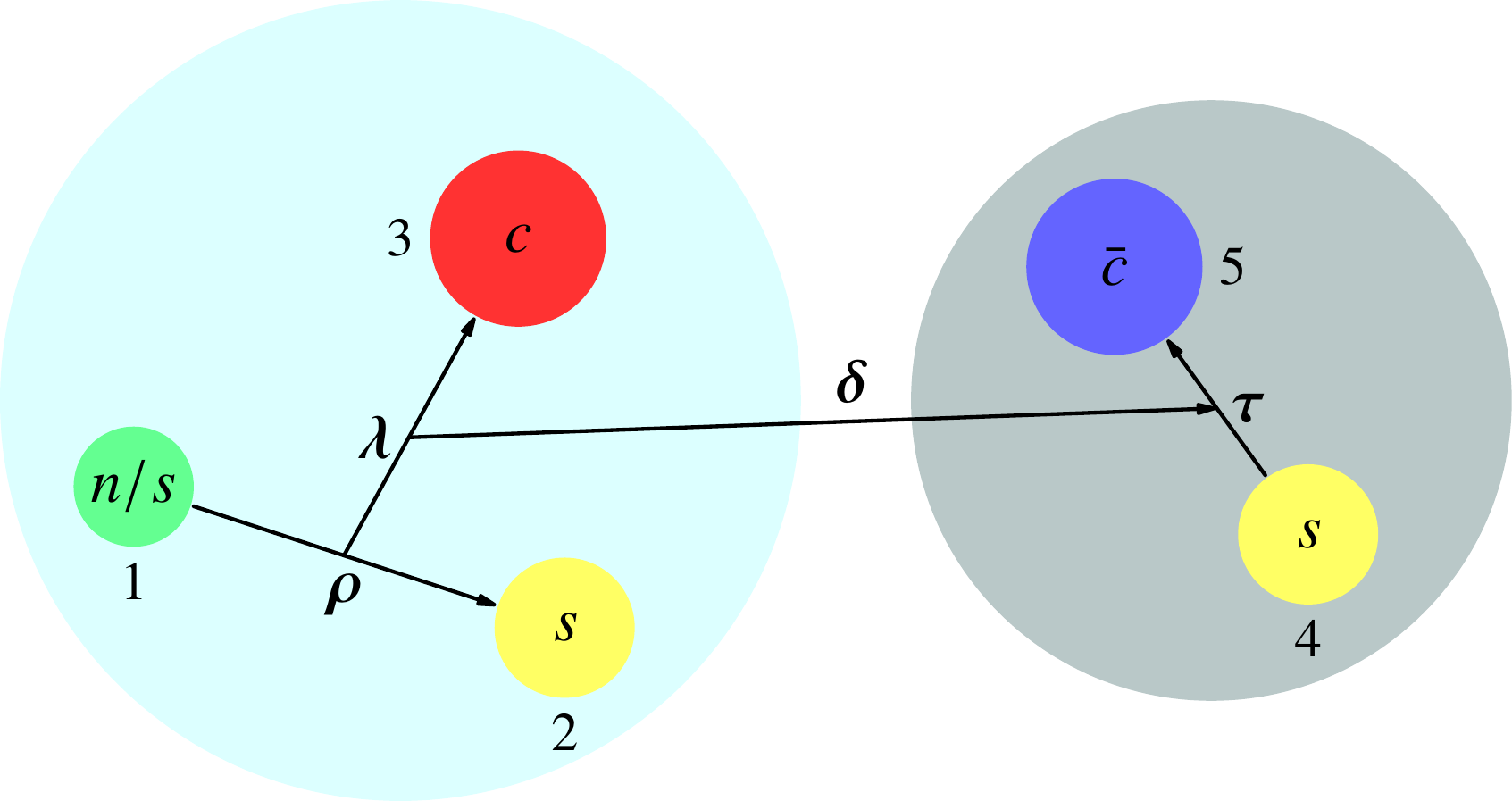}
\caption{The Jacobi coordinates of the $\Xi_c^{(\prime,*)} \bar{D}_s^{*}/\Omega_{c}^{(*)}\bar D_s^{*}$-type hadronic molecular states. Here, $n$ stands for the up quark or the down quark, $s$ represents the strange quark, and $c$ denotes the charm quark.}
\label{fig:HM}
\end{figure}

As depicted in Fig. \ref{fig:HM}, the spatial wave functions of the hadronic molecular states are examined using the Jacobi coordinates ${\bf R}={{\bm \rho},{\bm \lambda},{\bm \tau},{\bm \delta}}$. These coordinates are employed to define the spatial distributions of the respective wave functions
\begin{eqnarray}
{\bm \rho}&=&{\bf r}_2-{\bf r}_1,\nonumber\\
{\bm \lambda}&=&{\bf r}_3-\frac{m_1{\bf r}_1+m_2{\bf r}_2}{m_1+m_2},\nonumber\\
{\bm \tau}&=&{\bf r}_5-{\bf r}_4,\nonumber\\
{\bm \delta}&=&\frac{m_4{\bf r}_4+m_5{\bf r}_5}{m_4+m_5}-\frac{m_1{\bf r}_1+m_2{\bf r}_2+m_3{\bf r}_3}{m_1+m_2+m_3},
\end{eqnarray}
and the spatial wave functions of the initial and final hadron states can be explicitly expressed as $\phi({\bm \rho})\phi({\bm \lambda})\phi({\bm \tau})\phi({\bm \delta})$. On the other hand, the coordinates of the quarks ${\bf r}_j~(j=1-5)$ can be written as
\begin{eqnarray}
{\bf r}_j=\sum\limits_n\alpha_{jn}{\bf R}_n=\alpha_{j1}{\bm \rho}+\alpha_{j2}{\bm \lambda}+\alpha_{j3}{\bm \tau}+\alpha_{j4}{\bm \delta}.
\end{eqnarray}
Thus, the ${\bf q}\cdot{\bf r}_j$ can be expanded by
\begin{equation}
{\bf q}\cdot{\bf r}_j={\alpha}_{j1}{\bf q}\cdot{\bm \rho}+{\alpha}_{j2}{\bf q}\cdot{\bm \lambda}+{\alpha}_{j3}{\bf q}\cdot{\bm \tau}+{\alpha}_{j4}{\bf q}\cdot{\bm \delta}.
\end{equation}
For the sake of convenience, we utilize a matrix representation to present the variable $\alpha$, which is defined as
\begin{equation}
\alpha=\left(
\begin{array}{cccc}
 -\frac{m_2}{m_1+m_2} & -\frac{m_3}{m_1+m_2+m_3} & 0 & -\frac{m_4+m_5}{m_1+m_2+m_3+m_4+m_5} \\
 \frac{m_1}{m_1+m_2} & -\frac{m_3}{m_1+m_2+m_3} & 0 & -\frac{m_4+m_5}{m_1+m_2+m_3+m_4+m_5} \\
 0 & \frac{m_1+m_2}{m_1+m_2+m_3} & 0 & -\frac{m_4+m_5}{m_1+m_2+m_3+m_4+m_5} \\
 0 & 0 & -\frac{m_5}{m_4+m_5} & \frac{m_1+m_2+m_3}{m_1+m_2+m_3+m_4+m_5} \\
 0 & 0 & \frac{m_4}{m_4+m_5} & \frac{m_1+m_2+m_3}{m_1+m_2+m_3+m_4+m_5} \\
\end{array}
\right).
\end{equation}
To describe the spatial wave functions of the baryons and mesons, we take the simple harmonic oscillator (SHO) wave function, i.e.,
\begin{eqnarray}\label{eq:sho}
\phi_{n,l,m}(\beta,{\bf r})&=&\sqrt{\frac{2n!}{\Gamma(n+l+\frac{3}{2})}}L_{n}^{l+\frac{1}{2}}(\beta^2r^2)\beta^{l+\frac{3}{2}}\nonumber\\
&&\times {\mathrm e}^{-\frac{\beta^2r^2}{2}}r^l Y_{l m}(\Omega),
\end{eqnarray}
where $Y_{l m}(\Omega)$ is the spherical harmonic function, $L_{n}^{l+\frac{1}{2}}(x)$ is the associated Laguerre polynomial, while $n$, $l$, and $m$ are the radial, orbital, and magnetic quantum numbers of the hadron, respectively. The parameter $\beta$ in Eq. (\ref{eq:sho}) corresponds to the SHO wave function and has been approximately estimated to be around $0.4~{\rm GeV}$ in previous theoretical studies \cite{Liu:2011yp, Li:2008xy}. In this work, we adopt this value for consistency. However, it is worth noting that the transition magnetic moment and the radiative decay width may exhibit slight variations when scanning the $\beta$ value within the range of $0.3~{\rm GeV}$ to $0.5~{\rm GeV}$. For the hadronic molecular state, which represents a loosely bound system, the spatial wave function (the ${\bm \delta}$-degree of freedom in Fig. \ref{fig:HM}) significantly deviates from the SHO wave function. Hence, we derive the accurate spatial wave function for the molecular state by quantitatively solving the Schr\"{o}dinger equation.

Furthermore, we employ the following equations \cite{Khersonskii:1988krb} to expand the spatial wave function of the emitted photon, denoted as $e^{-i{\bf q}\cdot{\bf r}_j}$
\begin{eqnarray}
e^{-i{\bf q}\cdot{\bf r}_j}&=&e^{-i\sum\limits_n\alpha_{jn}{\bf q}\cdot{\bf R}_n}=\prod\limits_ne^{-i \alpha_{jn}{\bf q}\cdot{\bf R}_n},\\
e^{-i\alpha_{jn}{\bf q}\cdot{\bf R}_n}&=&\sum\limits_{l=0}^\infty\sum\limits_{m=-l}^l4\pi(-i)^lj_l(\alpha_{jn}qR_n)\nonumber\\
&&\times Y_{lm}^*(\Omega_{\bf q})Y_{lm}(\Omega_{{\bf R}_n}).
\end{eqnarray}
Here, $j_l(x)$ is the spherical Bessel function. In the $S$-wave scheme, there only exists the $l=0$ part, which leads to the following relation
\begin{eqnarray}
&&\langle \phi_{0,0,0}(\beta^\prime,{\bf R}_n)|e^{-i \alpha_{jn}{\bf q}\cdot{\bf R}_n}|\phi_{0,0,0}(\beta,{\bf R}_n)\rangle\nonumber\\
&&=\int \phi_{0,0,0}(\beta^\prime,{\bf R}_n)j_0(\alpha_{jn}qR_n)\phi_{0,0,0}(\beta,{\bf R}_n){\rm d}^3 {\bf R}_n\nonumber\\
&&=\frac{2\sqrt{2}(\beta^\prime\beta)^{3/2}}{(\beta^{\prime2}+\beta^2)^{3/2}}e^{-\frac{\alpha_{jn}^2q^2}{2(\beta^{\prime2}+\beta^2)}}.
\end{eqnarray}
Based on the above discussion, we can perform a comprehensive analysis of the contributions from the spatial wave functions of the emitted photon, baryons, mesons, and hadronic molecules to the transition magnetic moment and the radiative decay width.

\end{document}